\documentclass[iop]{emulateapj} 
\usepackage{apjfonts}
\usepackage{psfig}
\usepackage{mathrsfs}
\usepackage{amsmath}

\newcommand{\rmxaa}{Rev. Mexicana Astron. Astrofis. } 
\newcommand{\RRab}{RR{\em ab}}
\newcommand{\RRc}{RR{\em c}}
\slugcomment{AJ, submitted (revised version, \today)}
\shorttitle{Time-Series Photometry for M72}
\shortauthors{Amigo et al.}

\begin{document}

\title{Time-Series $BVI$ Photometry for the Globular Cluster NGC~6981 (M72)\footnotemark[1,]\footnotemark[2,]\footnotemark[3]}

\footnotetext[1]{This research draws upon data distributed by the NOAO Science
Archive. NOAO is operated by the Association of Universities for
Research in Astronomy (AURA) under cooperative agreement with the
National Science Foundation.}

\footnotetext[2]{Based in part on data obtained from the ESO Science Archive Facility.}

\footnotetext[3]{This paper makes use of data obtained from the Isaac Newton Group Archive
which is maintained as part of the CASU Astronomical Data Centre at the
Institute of Astronomy, Cambridge.}


\author{P. Amigo,\altaffilmark{4,5,6} P. B. Stetson,\altaffilmark{7}  M. Catelan,\altaffilmark{4,5} 
        M. Zoccali,\altaffilmark{4,5} \and H. A. Smith\altaffilmark{8}}

 \altaffiltext{4}{Pontificia Universidad Cat\'olica de Chile, Instituto de  
       Astrof\'\i sica, Av. Vicu\~{n}a Mackenna 4860, 
       782-0436 Macul, Santiago, Chile; e-mail: mcatelan, mzoccali@astro.puc.cl}

 \altaffiltext{5}{The Milky Way Millennium Nucleus, Av. Vicu\~{n}a Mackenna 4860, 782-0436 Macul, Santiago, Chile}

 \altaffiltext{6}{Universidad de Valpara\'iso, Departamento de F\'isica y Astronom\'ia, Gran Bretan\~na 1111, Playa Ancha, Valpara\'iso;
 e-mail: pia.amigo@dfa.uv.cl}

 \altaffiltext{7}{Dominion Astrophysical Observatory, Herzberg Institute of Astrophysics, National Research Council, 5071 West Saanich Road, Victoria, BC V9E 2E7, Canada;  
e-mail: Peter.Stetson@nrc-cnrc.gc.ca}

 \altaffiltext{8}{Department of Physics and Astronomy, Michigan State University, East Lansing, MI 48824; e-mail: smith@pa.msu.edu}

%

\begin{abstract}
We present new $BVI$ photometry of the globular cluster \objectname{NGC~6981} (M72), based mostly on ground-based CCD
archive images. We present a new color-magnitude diagram (CMD) that reaches almost four magnitudes below the turn-off level. We performed
new derivations of metallicity and morphological parameters of the evolved sequences, in good agreement with previous authors, obtaining a value
of ${\rm [Fe/H]} \simeq -1.50$ in the new UVES scale. We also identify the cluster's blue straggler population. Comparing the radial 
distribution of these stars with the red giant branch population, we find that the blue stragglers are more centrally concentrated, 
as found in previous studies of blue stragglers in globular clusters. Taking advantage of the large field of view covered by our
study, we analyzed the surface density profile of the cluster, finding extratidal main sequence stars out to $r \approx 14.1\arcmin$, 
or about twice the tidal radius. 
We speculate that this may be due to tidal disruption in the course of M72's orbit, in which case tidal tails associated with the 
cluster may exist. We also take a fresh look at the variable stars in the cluster, recovering all previous known variables, including 
three SX Phoenicis stars, and adding three previously unknown RR Lyrae (1 c-type and 2 ab-type) to the total census. Finally, comparing 
our CMD with unpublished data for M3 (NGC~5272), a cluster with similar metallicity and horizontal branch morphology, we found that both 
objects are essentially coeval.
\end{abstract}


\section{Introduction}
\label{sec:intro}

Galactic globular clusters (GGCs) represent a key ingredient for the understanding of the evolution of the Galaxy, both from a chemical 
and a stellar populations perspective. In particular, they represent the oldest systems in the Milky Way, and their relative ages 
accordingly play an important role in constraining the Galaxy's formation timescale (e.g., Salaris \& Weiss 2002; Mar\'in-Franch et al. 
2009; Dotter et al. 2010). A major advantage encountered in the study of GCs is the frequent presence of large RR Lyrae populations, 
which are of great interest due to several properties, including a relative narrow range in magnitude (which makes them excellent standard 
candles to determine distances) and also the presence of correlations between their pulsation properties and the properties of the GCs 
to which they belong (see, e.g., Zorotovic et al. 2010 and Cohen et al. 2011, for some recent examples). 

Another remarkable characteristic of the RR Lyrae stars in GCs is the Oosterhoff dichotomy (Oosterhoff 1939, 1944), which consists in 
the sharp division of the GGC system into two groups, according to the average period of their ab-type (i.e., fundamental-mode) RR Lyrae 
population: Oosterhoff type I (OoI), with $\langle P_{\rm ab} \rangle \approx 0.55$~d, and Oosterhoff type II (OoII), with 
$\langle P_{\rm ab} \rangle \approx 0.65$~d. This dichotomous behavior applies exclusive to Galactic globulars, since nearby extragalactic 
systems somehow preferentially occupy a period range that is intermediate between groups OoI and OoII (Catelan 2009; Catelan et al. 
2012). In this sense, the Oosterhoff dichotomy provides important information that must be taken into account, when studying the early 
formation history of the Milky Way and its dwarf satellite galaxy system. In this context, obtaining a complete census of variability 
in GGCs is especially important, for it may help identify outliers in the Oosterhoff dichotomy context, and hence potentially point 
to GCs of extragalactic origin. 

Perhaps surprisingly, after more than a century of GGC variability studies, there is still a large number of GCs which lack 
time-series analyses, particularly using modern techniques. In this sense, the advent of CCD detectors, combined with sophisticated 
crowded-field photometry and image subtraction techniques (e.g., Stetson 1987, 1994; Alard \& Lupton 1998; Alard 2000; Bramich 2008) 
has proven of key importance for unveiling the presence of large numbers of previously unknown variable stars in GCs, particularly in 
their crowded cores.

 %
  \begin{figure*}[t]
  \centering
  \includegraphics[width=10cm]{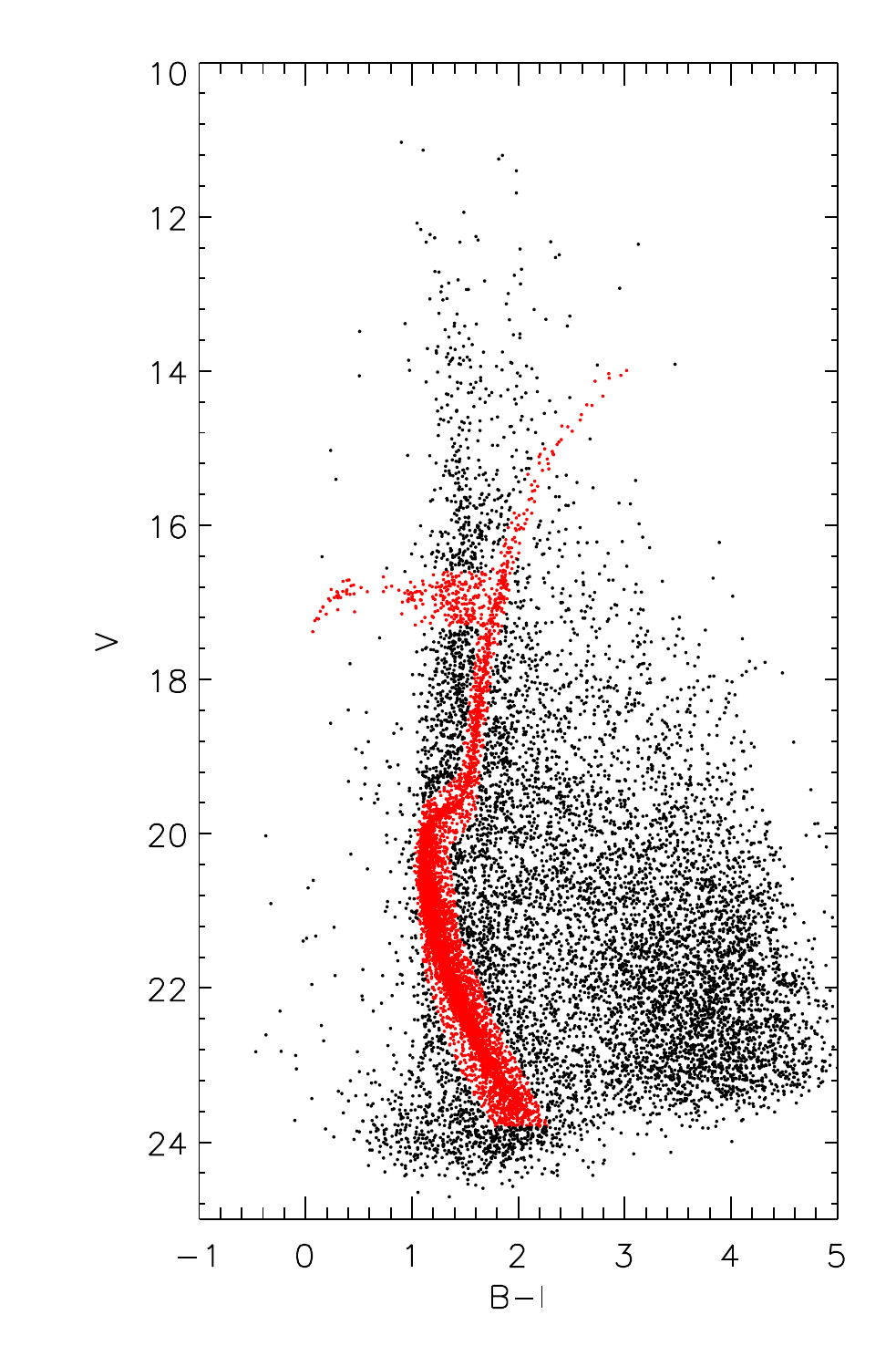}
  \caption{Acceptance ``box'' in the $B-I$, $V$ CMD ({\em red dots}). Stars within this ``box'' are considered to be cluster candidates 
  (plus field contaminants). The number of cluster stars outside the box ({\em black dots}) is assumed to be small.}\label{accbox}
  \end{figure*}
 %
 %

According to the Dec. 2010 version of the Harris (1996) catalog, the position of \objectname{NGC~6981}  is $\rm \alpha = 20^h53^m27^s.9$,
$\delta = -12\degr 32\arcmin 13 \arcsec$ (J2000), or $\ell = 35\fdg16$, $b = -32\fdg68$, with a low interstellar extinction of $E(\bv) = 0.05$.
Very recently, Bramich et al. (2011) published the results of time-series imaging of this cluster, performed with difference imaging, and 
updated the cluster's variable star census. In the framework of our long-term project to complete the census of variable stars in GGCs
(Catelan et al. 2006), and with the advantage of a longer time-baseline based on archival and new images, we have revisited the variability 
content of \objectname{NGC~6981}. Importantly, with the large number of images used and a large field of view (FOV), we were able to obtain a 
much deeper, spatially resolved, high-precision CMD, which allowed us to study several properties of the cluster, including its surface
brightness profile and the presence of extratidal stars. 

This paper is organized as follows. In Section~\ref{sec:data} our data and data analysis techniques are described. In Section~\ref{sec:extratidal}
we analyze the structural parameters of the cluster. Section~\ref{sec:cmd} provides our analysis of the different branches of the CMD.
In Section~\ref{sec:age}, we compare our results with another well-studied GC of similar metallicity, namely M3 (NGC~5272), in order to 
position M72 in a relative age scale, compared with other GGCs. In Section~\ref{sec:variables}, we discuss our findings regarding the 
cluster's variable star population. We close in Section~\ref{sec:conclu} with a summary of our results.

\begin{table*}[t]
\begin{center}
\footnotesize
\caption{\footnotesize{NGC 6981 Data Used in This Study.}}
\begin{tabular}{lccccccccccc}
\tableline
\tableline
Note   & Run ID   &  Dates             & Telescope    & Camera/Detector       & $U$ &  $B$ &  $V$ &  $R$ &  $I$ & 51 \\
\tableline
  1 & cmr      & 1986 Oct 28        & INT 2.5m     & GEC                   & - &  3 &  3 &  2 &  3 &  - \\
  2 & full2    & 1989 Aug 01-02     & CTIO 0.9m    & TI1                   & - &  6 &  8 &  2 &  - &  - \\
  3 & ct94jun  & 1994 Jun 28-Jul 01 & CTIO 0.9m    & Tek2K$_4$               & - & 12 & 12 &  - & 12 &  - \\
  4 & ct94nov  & 1994 Nov 25-28     & CTIO 0.9m    & Tek2K$_4$               & - &  2 &  2 &  - &  2 &  - \\
  5 & ct94sep  & 1994 Sep 22-26     & CTIO 0.9m    & Tek2K$_3$               & - & 25 & 25 &  - & 24 &  - \\
  6 & ct95jun  & 1995 Jun 21-25     & CTIO 0.9m    & Tek2K$_3$               & - & 15 & 15 &  - & 15 &  - \\
  7 & bond8    & 1996 Sep 25        & KPNO 0.9m    & t2ka                  & 1 &  1 &  1 &  - &  1 &  - \\
  8 & dmd      & 1998 Jun 25        & JKT 1.0m     & TEK4                  & - &  - &  2 &  - &  2 &  - \\
  9 & int98    & 1998 Aug 20        & INT 2.5m     & WFC                   & - &  2 &  2 &  - &  2 &  - & $\times 4$ \\
10 & fors9906 & 1999 Jun 17-21     & ESO VLT 8.0m & FORS1                 & - &  - &  5 &  - & 24 &  - \\
11 & fors9908 & 1999 Aug 02        & ESO VLT 8.0m & FORS1                 & - &  - &  - &  - &  2 &  - \\
12 & fors9909 & 1999 Sep 02        & ESO VLT 8.0m & FORS1                 & - &  - & 19 &  - &  - &  - \\
13 & wfi5     & 2002 Jun 20        & ESO/MPI 2.2m & WFI                   & - &  4 &  4 &  - &  - &  - & $\times 8$ \\
14 & m72      & 2007 Apr 15-Aug 03 & CTIO 1.3m    & ANDICAM Fairchild 447 & - & 41 & 42 &  - & 41 &  - \\
15 & emmi8    & 2007 Jul 16        & ESO NTT 3.6m & EMMI MIT/LL mosaic    & - &  4 &  3 &  7 &  - &  - & $\times 2$ \\
16 & jul08    & 2008 Jul 27-29     & CTIO 4.0m    & Mosaic2               & - &  2 & 17 &  - & 17 &  - & $\times 8$ \\
17 & aug08    & 2008 Aug 26-27     & CTIO 4.0m    & Mosaic2               & 2 &  6 & 14 &  - & 16 &  3 & $\times 8$ \\
18 & int11    & 2011 Aug 22-Sep 24 & INT 2.5m     & WFC                   & 3 &  2 &  2 &  - &  2 &  - & $\times 4$ \\
\tableline
\end{tabular}
\label{tab:data}

\par \smallskip \scriptsize \textit{Notes}.
1. Observer ``CMR'' 
2. Observer L.~Fullton 
3. Observer A.~R.~Walker 
4. Observer A.~R.~Walker 
5. Observer A.~R.~Walker 
6. Observer A.~R.~Walker 
7. Observer H.~E.~Bond 
8. Observer ``DMD'' 
9. Observer unknown 
10. Program ID 63.L-0342(B); ESO internal PI-COI ID 403; observer unknown 
11. Program ID 63.L-0342(B); ESO internal PI-COI ID 403; observer unknown 
12. Program ID 63.L-0342(B); ESO internal PI-COI ID 403; observer unknown
13. Program ID 69.D-0582(A); ESO internal PI-COI ID 361; observer unknown
14. PI unknown; observers C.~Aguilera, J.~Espinoza, A.~Pasten
15. Program ID 59.A-9002(A); ESO internal PI-COI ID 50002; observer unknown
16. Proposal ID 2008A-289; observer A.~R.~Walker 
17. Proposal ID 2008B-155; observer A.~R.~Walker 
\end{center}
\end{table*}

\section{Observations}\label{sec:data}

The observational material for this study consists of 1,139
individual CCD images from 18 observing runs.  These images are
contained within a data archive maintained by PBS, and include~--
among others~-- images obtained specifically for this project. 
Notable among these is one series of exposures dating from 2007
April 7 through 2007 August 3:  observations were taken on 42 of
these 110 nights, with no gap greater than eight nights.  Summary
details of the 18 observing runs are given in Table 1. 
Considering all these images together, the median seeing for our
observations was 1.4 arcseconds; the 25th and 75th percentiles
were 1.2 and 1.6 arcseconds; the 10th and 90th percentiles were
1.0 and 1.8 arcseconds.  

The photometric reductions were all carried out by PBS using
standard DAOPHOT/ALLFRAME procedures (Stetson 1987, 1994) to
perform profile-fitting photometry, which was then referred to a
system of synthetic-aperture photometry by the method of growth-curve
analysis (Stetson 1990).  Calibration of these instrumental data
to the photometric system of Landolt (1992; see also Landolt 1973,
1983) was carried out as described by Stetson (2000, 2005).  If we
define a ``dataset'' as the accumulated observations from one CCD
chip on one night with photometric observing conditions, or one
chip on one or more nights during a single observing run with
non-photometric conditions, the data for M72 were contained within
83 different datasets, each of which was individually calibrated
to the Landolt system.  Of these 83 datasets, 60 were obtained and
calibrated as ``photometric,'' meaning that photometric zero
points, color transformations, and extinction corrections were
derived from all standard fields observed during the course of
each night, and were applied to the M72 observations.  The other 23
datasets were reduced as "non-photometric"; in this case, color
transformations were derived from all the standard fields
observed during the course of the observing run, but the
photometric zero point of each individual CCD image was derived
from local M72 photometric standards contained within the image
itself.  

The different cameras employed projected to different areas on the
sky, and of course the telescope pointings differed among the
various exposures.  The WFI, WFC, Mosaic2, and EMMI imagers, in particular,
consist of mosaics of non-overlapping CCD detectors.  Therefore,
although we have 1,139 images, clearly no individual star appears
in all those images.  In fact, no star appeared in more than 169
$B$-band images, 200 $V$ images, or 190 $I$ images.  Considering the
entire area surveyed, the {\it median\/} number of observations
for one of our stars is 13 in $B$, 34 in $V$, and 35 in $I$.  In the
immediate vicinity of the cluster (within a radius of order 7
arcminutes), however, most stars appeared in at least 100 images
in each of $B$, $V$, and $I$.  

There were insufficient $U$- and $R$-band images to define local
standards in the M72 field, so we have not calibrated the $U$ or
$R$ data and make no use of them here.  In addition, during the
2008 August observing run on the CTIO Blanco 4m telescope, three
exposures (24 individual CCD images) were obtained in a DDO51
filter.  This bandpass measures the strength of a molecular
band of MgH and the ``b'' triplet lines of neutral magnesium, and
can be used as a gravity/metallicity discriminator in very cool
stars to distinguish likely member giants from foreground field
dwarfs.  While we do not employ the DDO51 data in the current
investigation, we may revisit these data in a future study. 
Still, even though we do not make use of the $U$, $R$, and DDO51
data in our photometric analysis, these images were included in
the ALLFRAME reductions for the additional information they give
on the completeness of the star list and the precision of the
astrometry.

\section{Extratidal Stars in NGC~6981}\label{sec:extratidal}
The extended FOV available in our study motivates a new derivation of the tidal radius for \objectname{NGC~6981}. Several recent papers  
in which the surface brightness distribution of GGCs was studied have revealed evidence of the presence of extratidal stars (e.g., 
Walker et al. 2011, hereafter W11; Correnti et al. 2011; Carballo-Bello et al. 2012; Salinas et al. 2012). This result is not entirely 
surprising, for two main reasons. First, the number of wide-field CCD observations of GGCs has increased dramatically since the classical 
studies dealing with the structural parameters of these objects were published (e.g., Grillmair et al. 1995). Second, clusters are expected 
to undergo tidal disruption in the course of their lives, due to their interactions with the Galaxy as they orbit in the halo 
\citep[e.g.,][]{fz01,bm03}, and direct evidence for this is being increasingly found in many GGCs \cite[e.g.,][]{moea01,moea03,gj06,jg10,ebea11}. 
Indeed, there is growing evidence that many GGCs may indeed have been much more massive in the past, some possibly being the remains of disrupted 
dwarf galaxies acting as building blocks in the hierarchical scenarios for galaxy formation \citep[e.g.,][and references therein]{tb08,vc11,smea12}. 

To study the structural parameters of \objectname{NGC~6981} and to find the level of contamination by field stars, we followed the method 
described in W11, which
is based on a similar dataset, and with a large FOV that extends more than 30\arcmin \ from the cluster center.
First, we identify an {\em acceptance box} in the $B-I$, $V$ plane, as shown in Figure~\ref{accbox}. We used the $B-I$, $V$ plane, as suggested
by W11, because of its stronger sensitivity to temperature, allowing a more robust determination of the ridgeline (more details on the ridgeline 
construction are given in Sect.~\ref{sec:cmd}). Also, the usage of all three filters available helps avoid spurious detections. 
  \begin{figure*}[t]
  \epsscale{1.0}
  \centering
  \includegraphics[width=10cm]{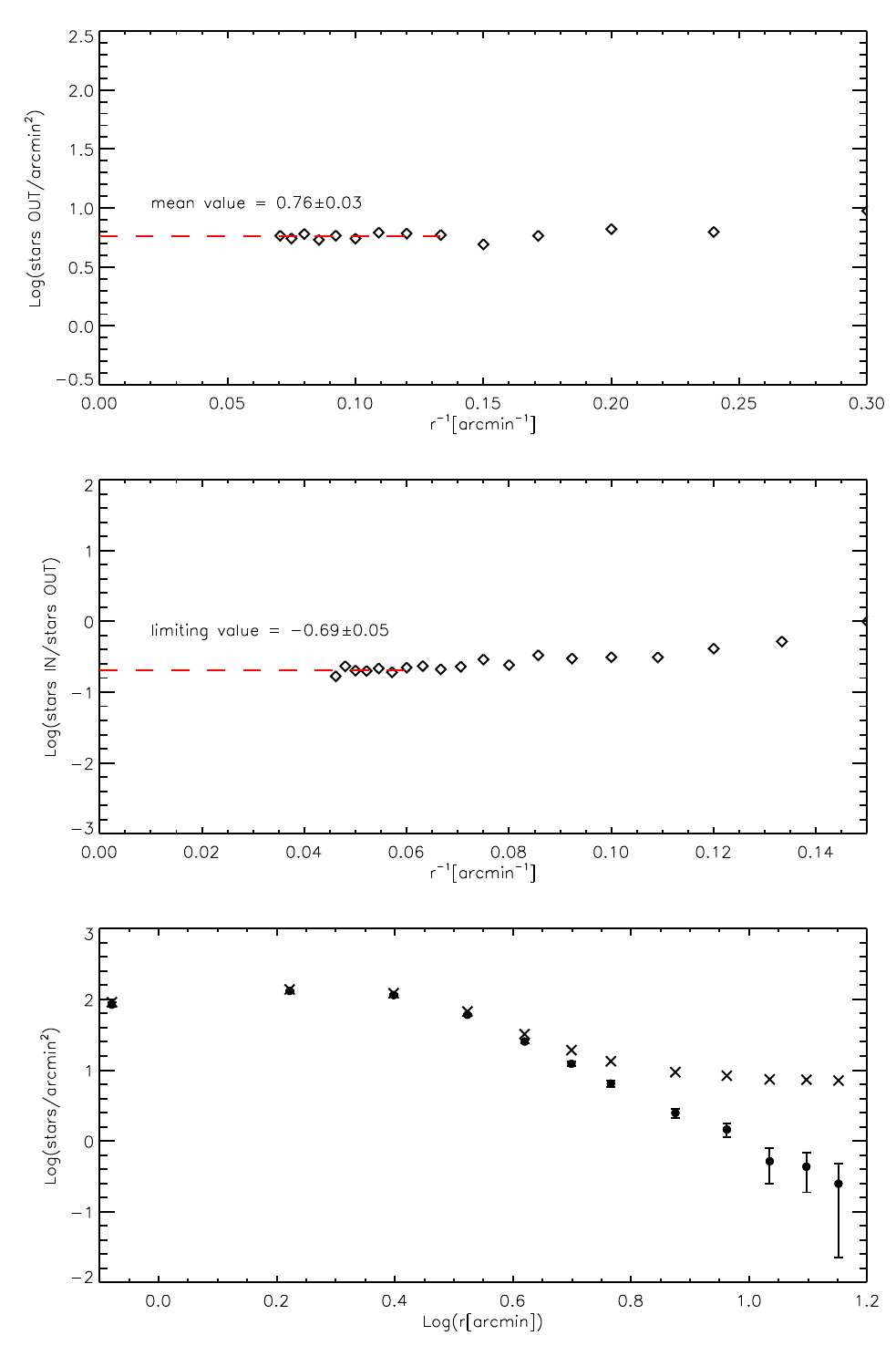}
  \caption{{\em Top}: logarithmic surface density as a function of the inverse of the radial distance, for stars outside the acceptance box.
  The mean value corresponds to the surface density of field stars outside the box of $\sim 5.8 \, {\rm stars/arcmin^2}$. {\em Middle}:
  ratio between the number of stars inside the box and those outside the box, extending the annuli in the entire FOV. 
  {\em Bottom} Surface density profile of NGC~6981, before ({\em crosses}) and after ({\em filled circles}) 
  subtraction of field stars. The error bars are Poissonian.}
 \label{surfdensfield}
  \end{figure*}
 %
 \begin{figure*}[htbp!]
  \centering
   \includegraphics[width=18cm]{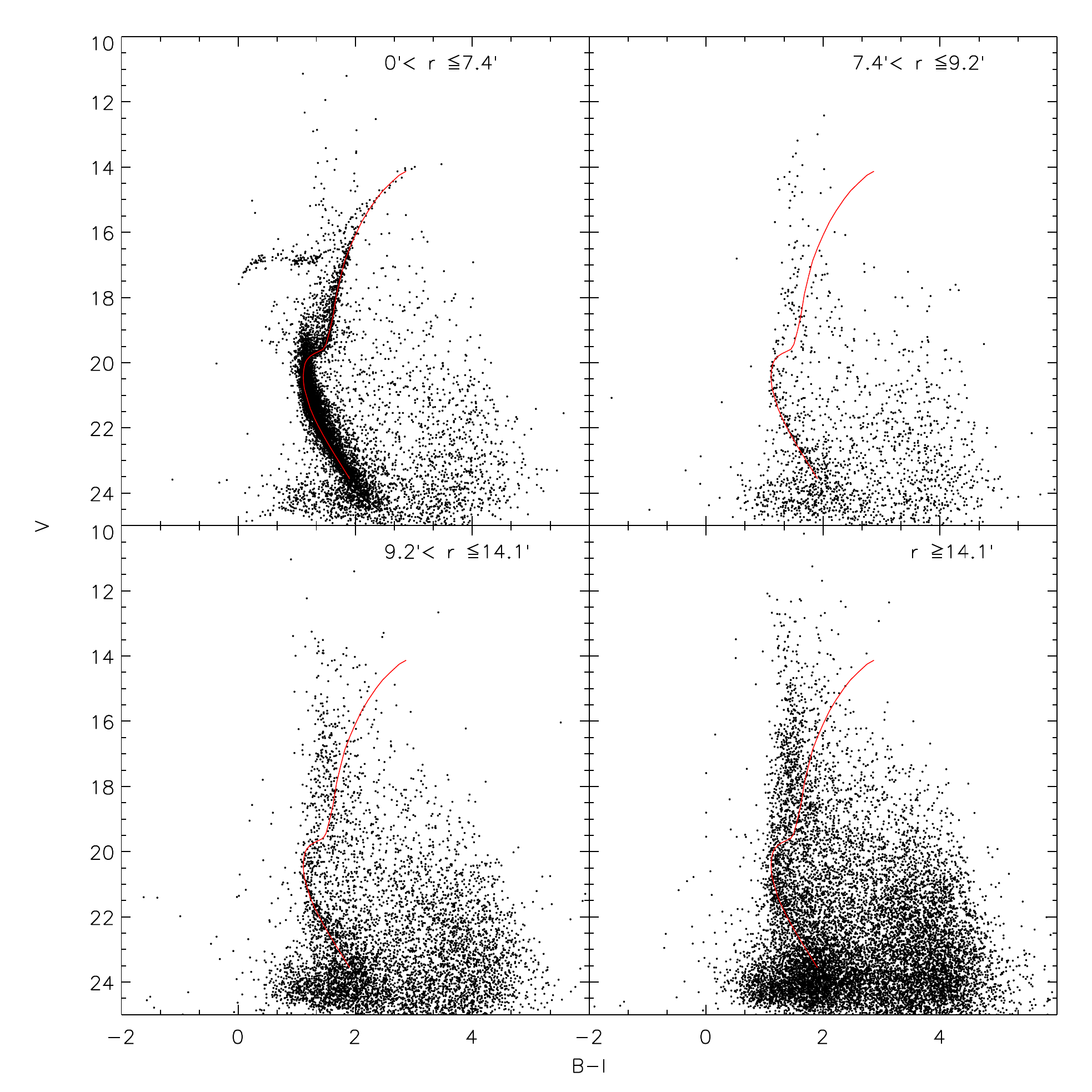}
   \caption{M72 CMD at different radial annuli. MS stars are distinguishable out to at least $r \approx 14\arcmin$.}
  \label{radii}
   \end{figure*}
 %
  %
  %
   \begin{figure}[htbp!]
   \centering
   \includegraphics[width=0.5\textwidth]{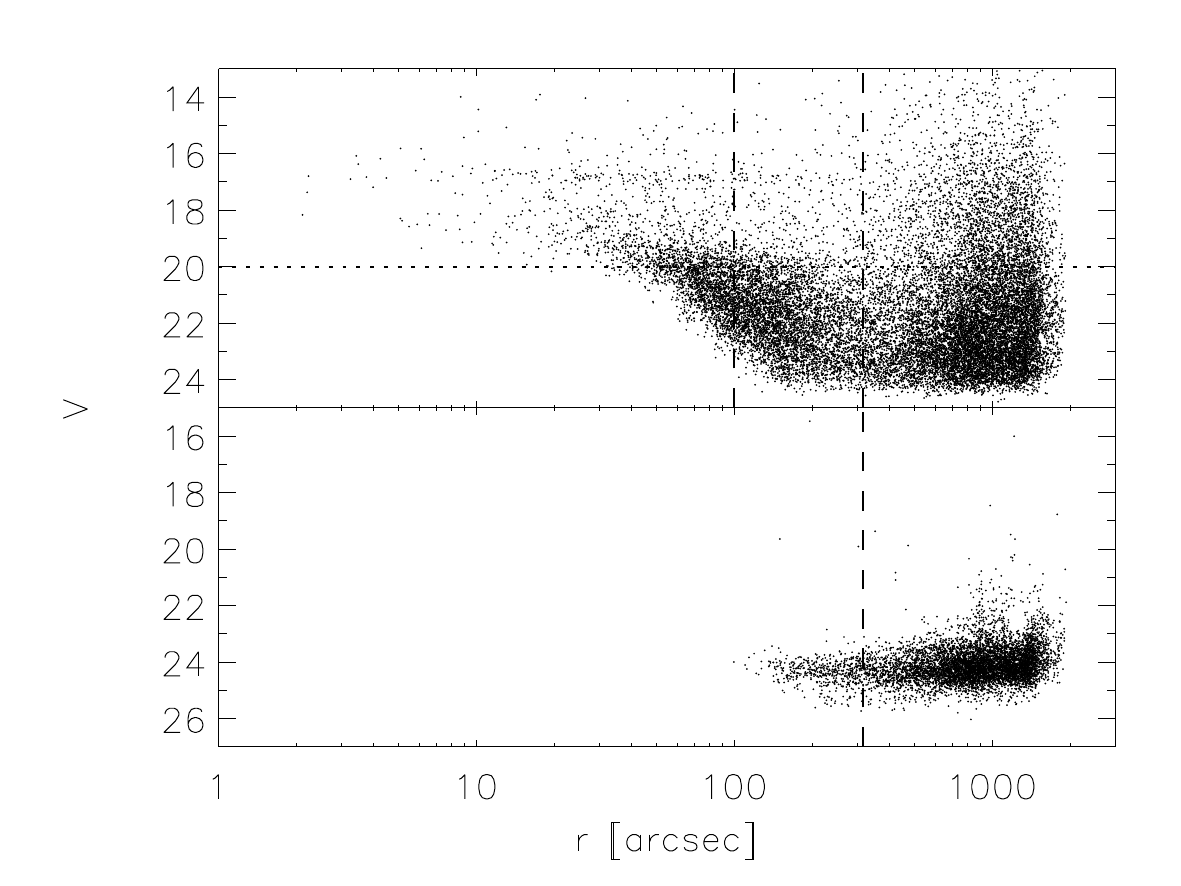}
   \caption{$V$ magnitude vs. radius for stars with $\sigma(B-I) \leq 0.1$ ({\em top}) and $0.1 \leq \sigma(B-I) \leq 0.3$ ({\em bottom}).
    More details in text.}
  \label{Vradius}
   \end{figure}
 
   \begin{figure*}[htbp!]
   \centering
   \includegraphics[width=18cm]{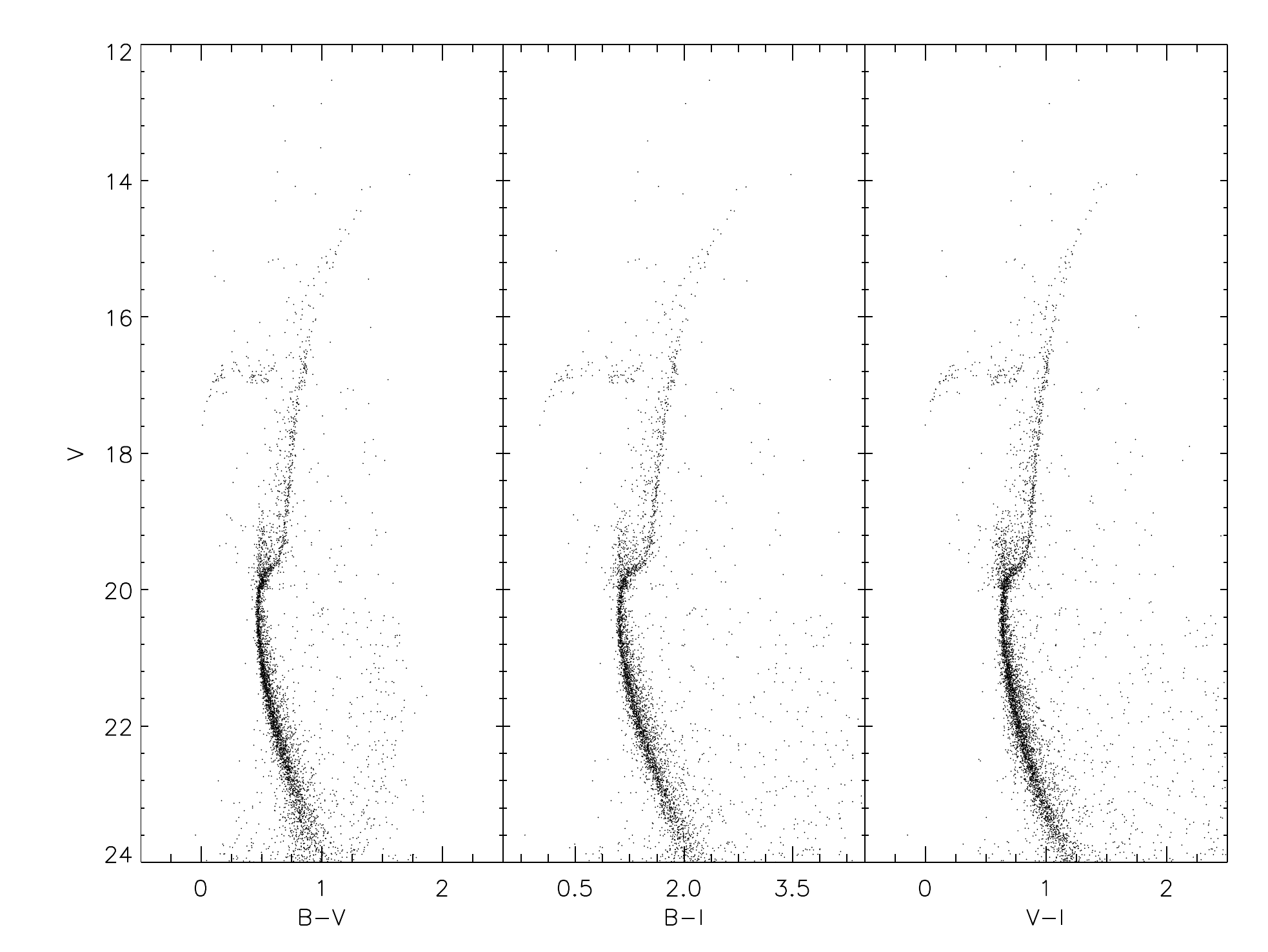}
   \caption{$BVI$ photometry for NGC~6981, in the $V$, $\bv$ plane ({\em left}), $V$, $B-I$ plane ({\em middle}), and $V$, $V-I$ plane 
            ({\em right}). Selection criteria according to Figure~\ref{Vradius} was applied (see text for details).}
  \label{BVI}
   \end{figure*}
   
 %
 %
   \begin{figure}[htbp!]
   \centering
   \includegraphics[width=0.5\textwidth]{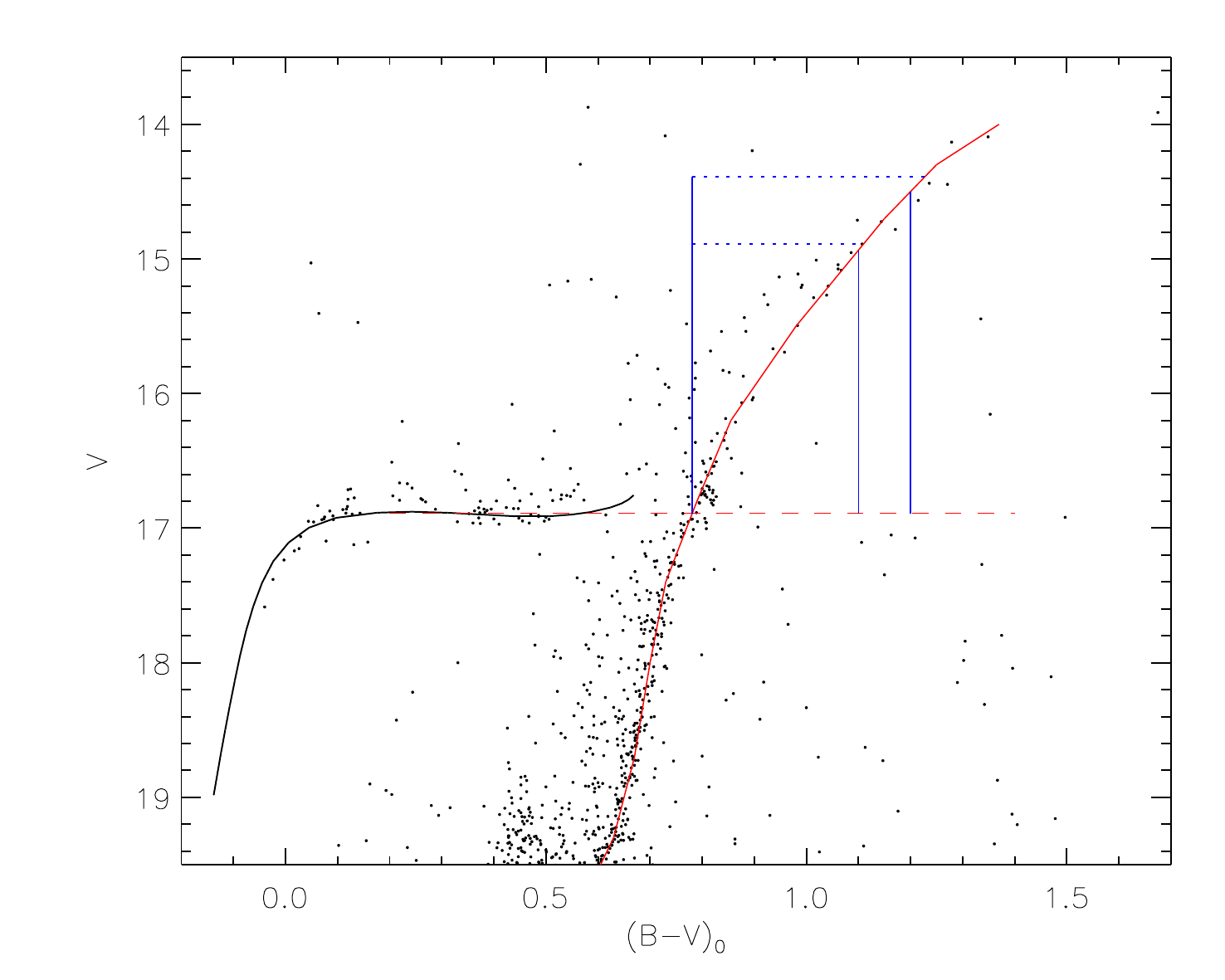}
    \caption{M72 CMD zoomed in around evolved sequences of the CMD. The solid black line corresponds to a Victoria-Regina model ZAHB 
             (VandenBerg et al. 2006) with $\rm [Fe/H] = -1.41$ and $\rm [\alpha/Fe] = +0.3$. The dashed line is the inferred ZAHB level, 
             at $V=$ 16.89 $\pm$ 0.02. The dashed horizontal lines  denote the levels 2.0 and 2.5~mag above the HB in $V$. 
    The solid vertical red line is defined by the color of the RGB at the HB level, $(\bv)_{g}$. 
    The solid vertical blue lines intersecting the ridgeline at the upper RGB indicate the  
    $\rm \Delta V$ measurements at $(\bv)_0 = 1.0$, 1.2, and 1.4 respectively. See Section~\ref{sec:metal} for more details}
  \label{ZAHB}
   \end{figure}
   
 %
  %
  %
   \begin{figure}[htbp!]
   \centering
    \includegraphics[angle=90,width=0.5\textwidth]{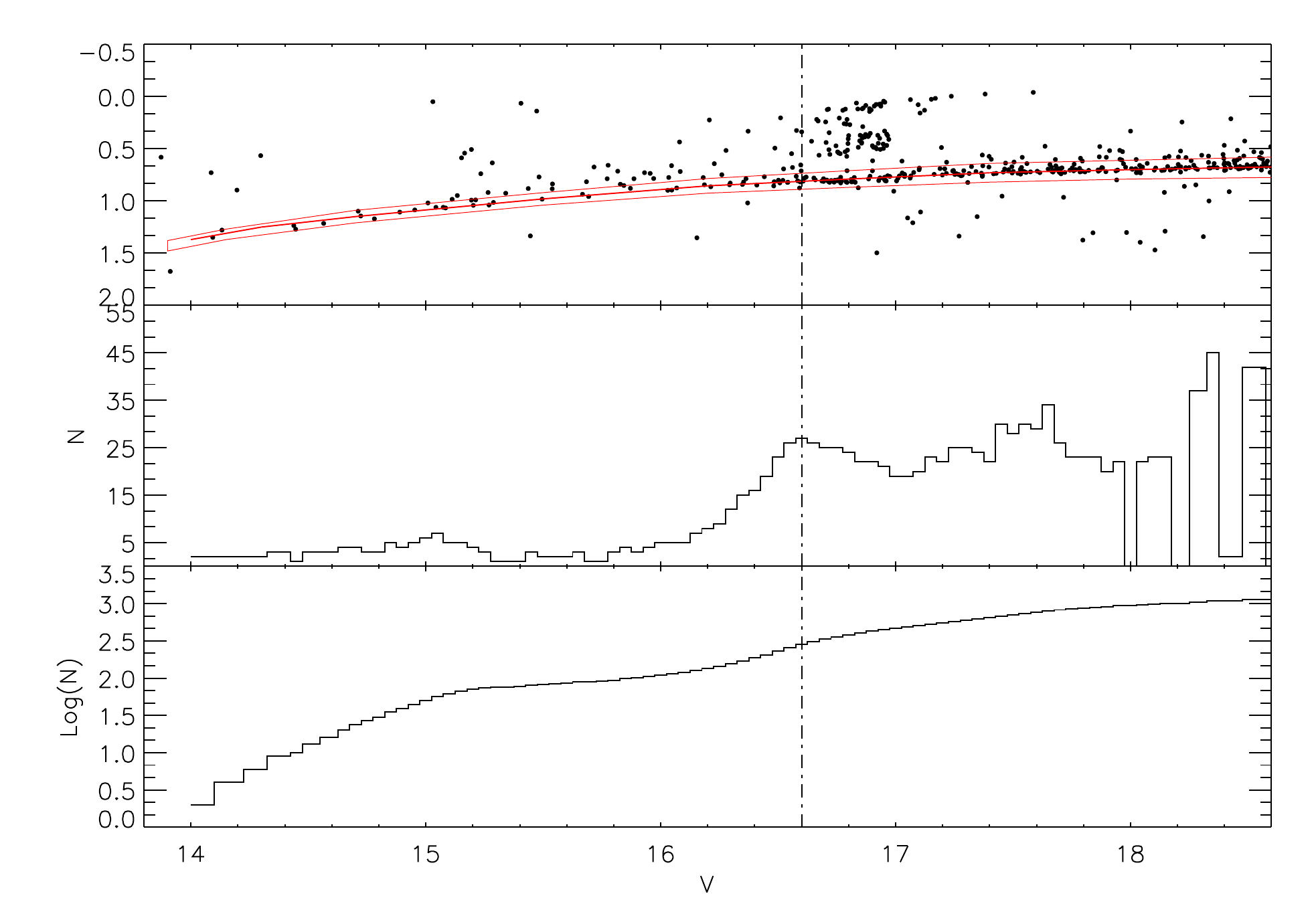}
    \caption{{\em Top}: Selection box of the RGB stars in the CMD. {\em Middle}: average-shifted histogram (differential luminosity function) 
    of the stars selected in the panel above. {\em Bottom}: cumulative luminosity function, showing the change 
    in slope brought about by the RGB bump. The dot-dashed line in the three panels marks the position of the bump, at $V_{\rm bump} =16.6$. }
   \label{historgb}
   \end{figure}
   
  %
  %
   \begin{figure}[htbp!]
   \centering
    \includegraphics[width=0.5\textwidth]{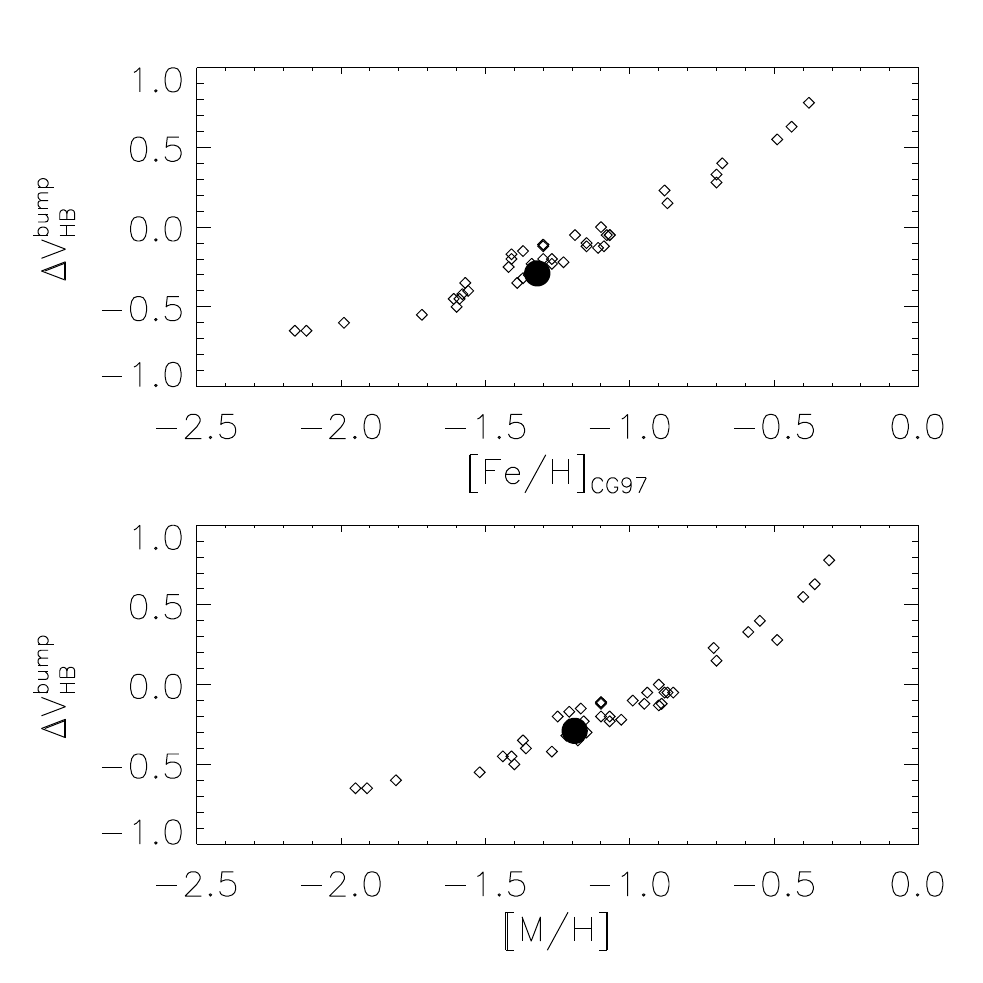}
    \caption{Parameter $\Delta V^{\rm bump}_{\rm HB}$ as a function of $\rm [Fe/H]_{CG97}$ ({\em top}) and $\rm [M/H]_{CG97}$ ({\em bottom}). 
             The diamonds correspond to data taken from F99, and the solid circle corresponds to the position of NGC~6981 measured in this study.}
   \label{bumpFeH}
   \end{figure}
 %
  \begin{figure}[htbp!]
   \centering
    \includegraphics[angle=90,width=0.5\textwidth]{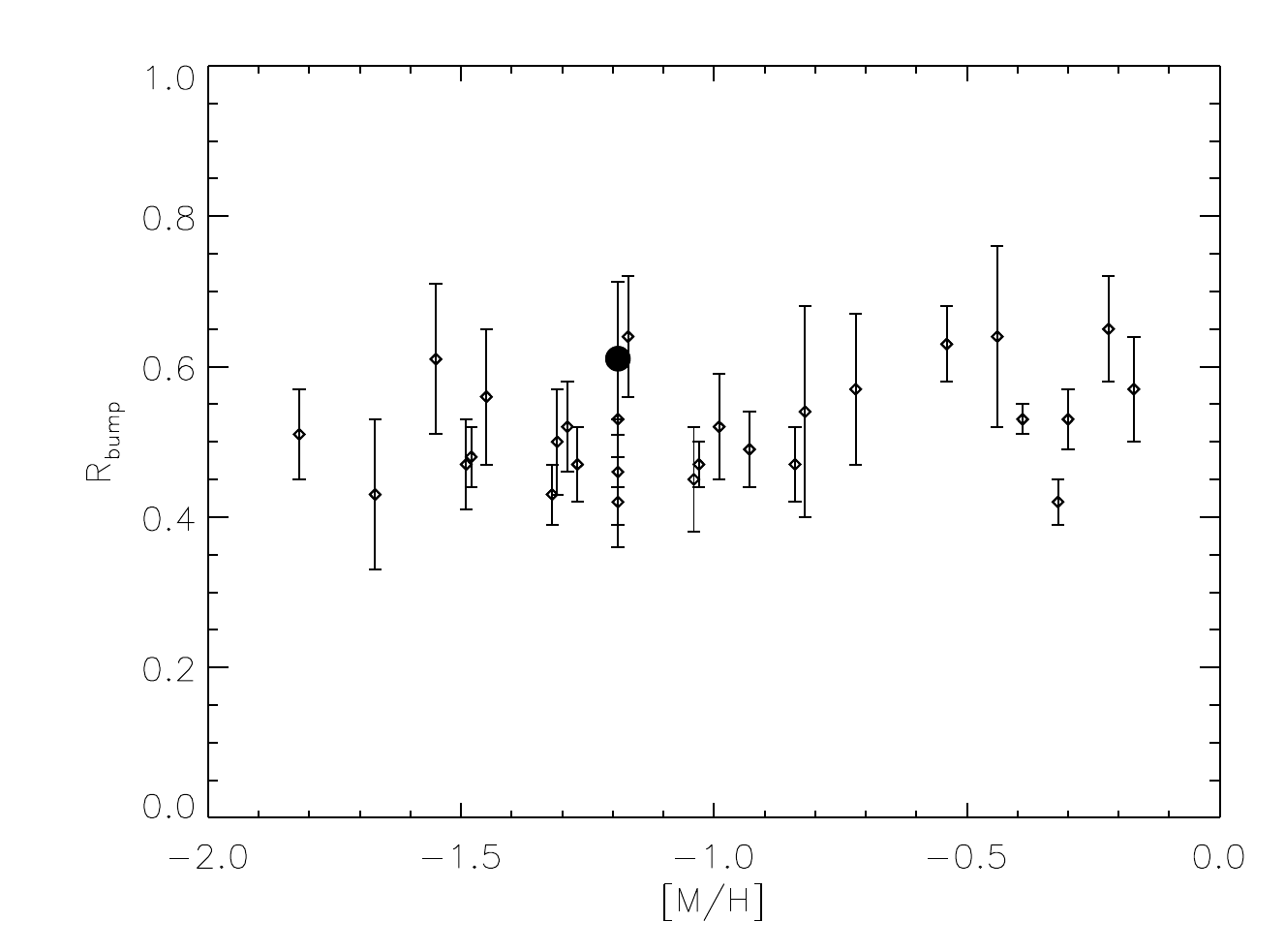}
    \caption{Parameter $R_{\rm bump}$ as a function of global metallicity. Data taken from Bono et al. (2001) ({\em empty diamonds}). 
    The {\em filled circle} indicates the value derived for M72 in our study.}
   \label{Rbump}
   \end{figure}
 %
   \begin{figure*}[t]
   \centering
    \includegraphics[width=18cm]{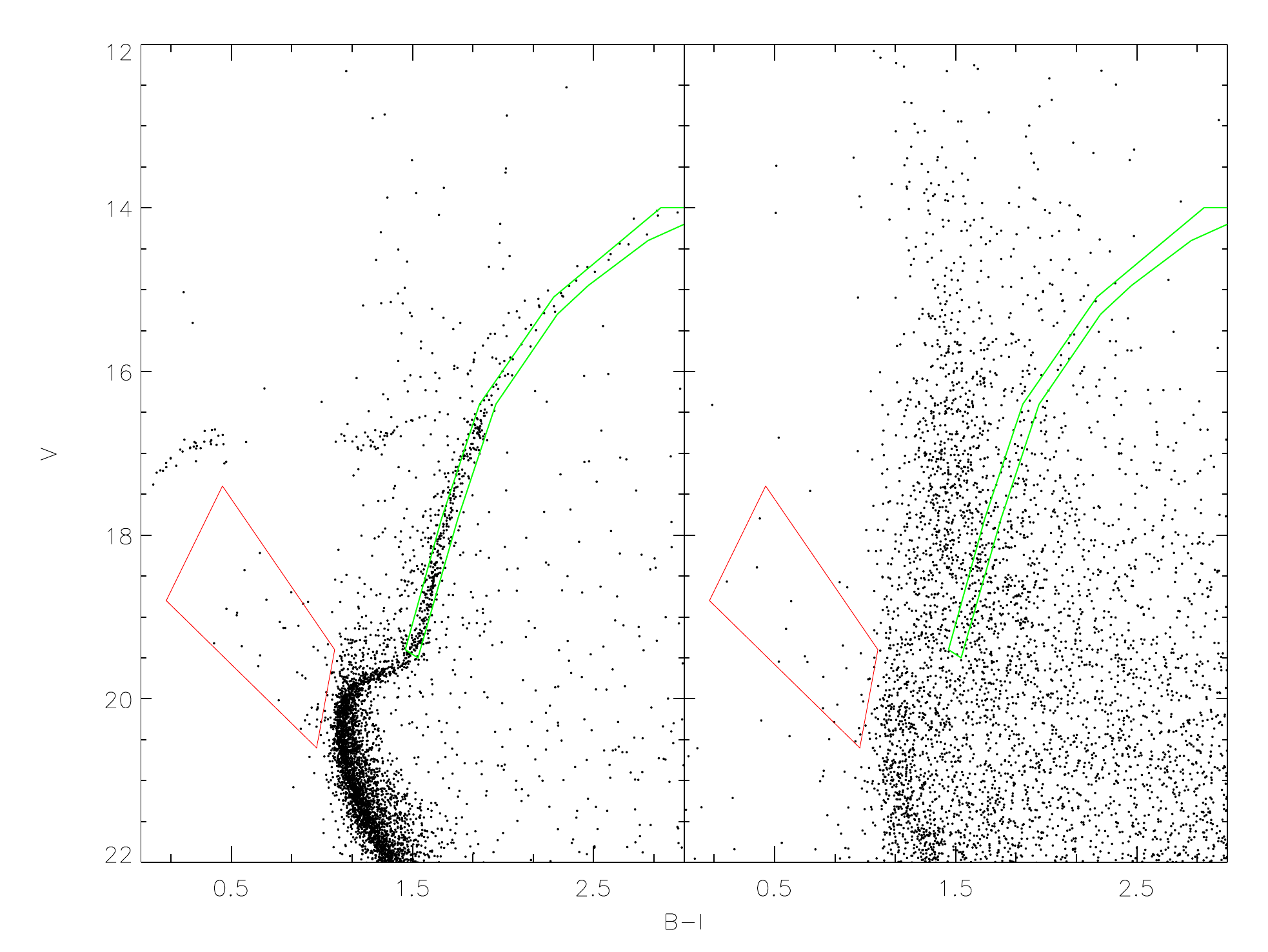}
   \caption{{\em Left}: Selection box for BSS ({\em red}) and RGB ({\em green}) populations, within $r \leq 7.1\arcmin$. {\em Right}: Stars
   with $r \geq 7.1\arcmin$, with the same selection boxes as in the other panel overplotted. }
   \label{BSScmd}
   \end{figure*}
 %
 %
   \begin{figure}[htbp!]
   \centering
   \includegraphics[width=0.5\textwidth]{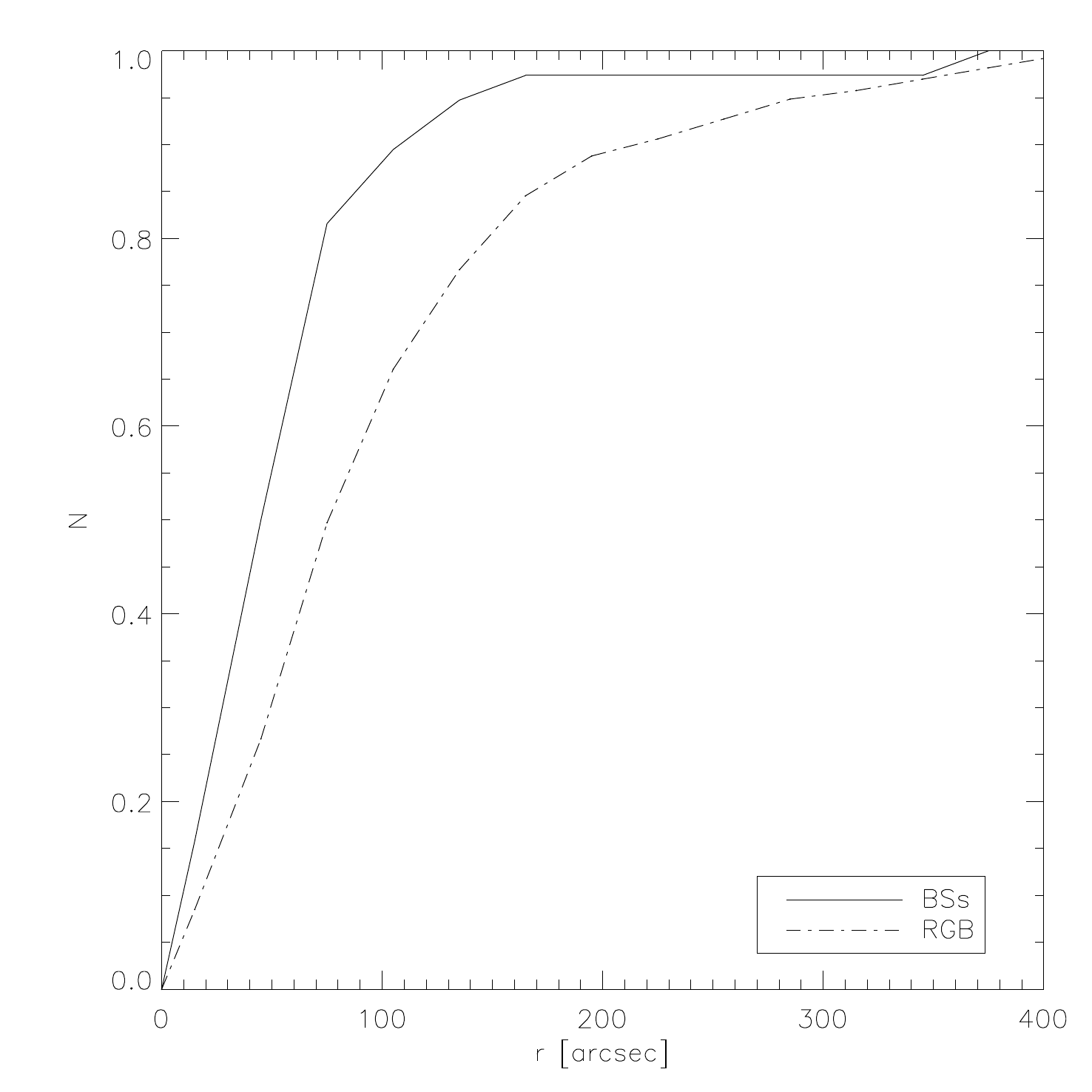}
    \caption{Radial cumulative distribution for both RGB and BSS populations in NGC~6981.}
  \label{BSScum}
   \end{figure}
   
 %
   \begin{figure}[htbp!]
  \centering
  \includegraphics[width=0.5\textwidth]{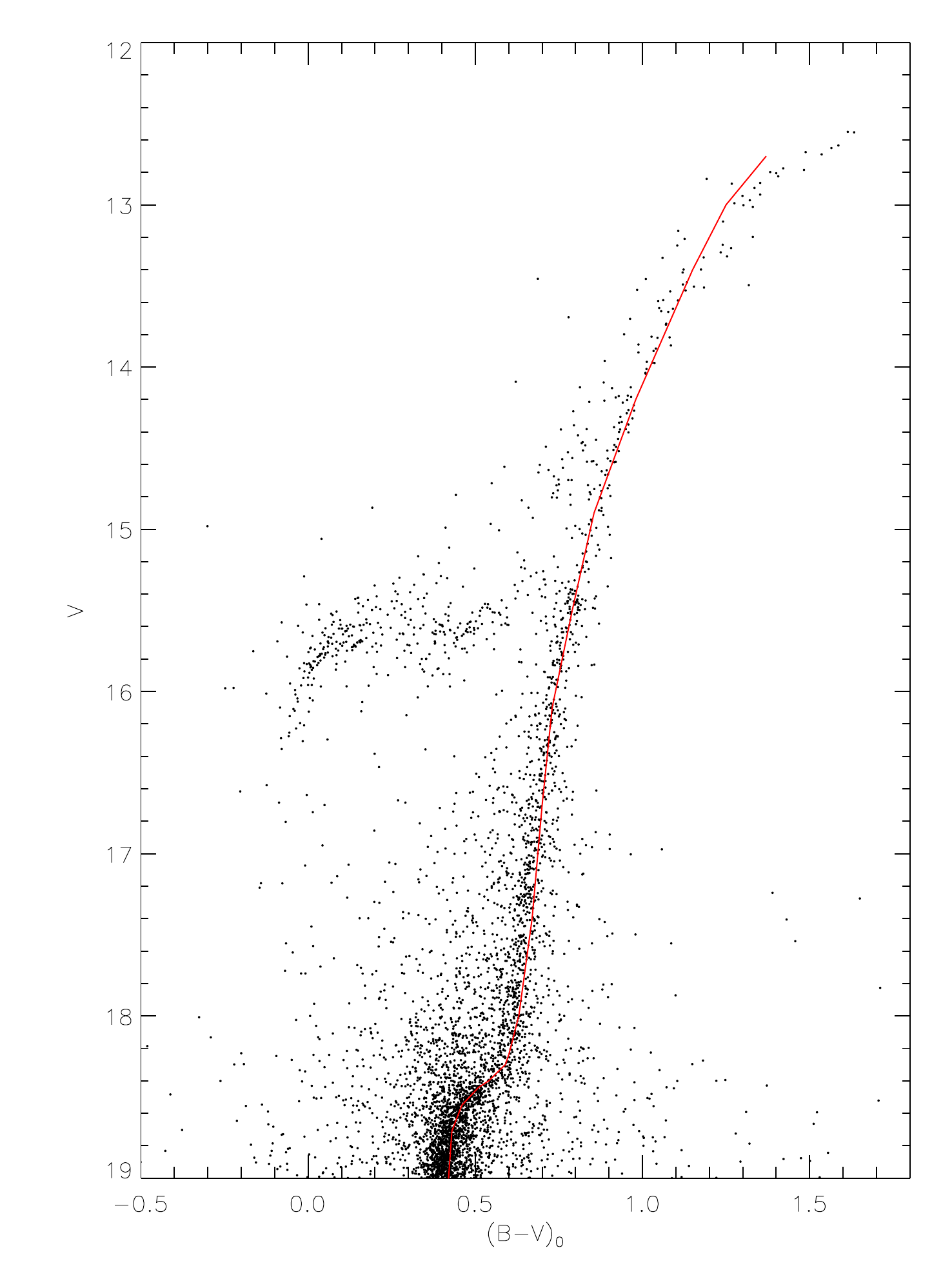}
     \caption{Comparison between our photometry for M3 ({\em points}) and the ridgeline of NGC~6981
     ({\em solid line}).}
    \label{compM3}
    \end{figure}
 %

The selected box is assumed to contain most true cluster members, and so the number of bona-fide members falling outside this box is assumed 
to be small. 
The contamination from field stars inside the box is small but not negligible. We did not include in the acceptance box the blue straggler 
star (BSS) population,  since they will have a very small effect in the estimation of overall number counts in the cluster, and certainly will
not affect the outer radii, due to its mostly central concentration. More details on the BSS population are given in Section~\ref{sec:BSS}.

Once the acceptance box has been defined, 
the field is divided in concentric annuli, each with a radius of 50\arcsec, in order to study the radial density 
profile inside and outside the acceptance box. Since the cluster is not centered in the FOV, the limit radius considered to be complete 
is the greatest circle within the FOV, this is $r \approx 14.1 \arcmin$, which corresponds to the edge closer to the center of the cluster. 

In Figure~\ref{surfdensfield} ({\em top panel}) is plotted the logarithmic surface density of all stars outside the acceptance box, as
a function of the inverse of the radial distance. Based on the
assumption that in the outermost annuli the presence of true
cluster members is negligible, it follows that the mean
density value of the last points provides a reasonable estimate of the surface density of field stars
outside the acceptance box. This leads to an estimated field contribution of $5.8 \, {\rm stars/arcmin^2}$. 
We then calculated the fraction of stars inside and outside the box. In this case, the fraction of counts in each case is not affected by 
incompleteness of the sample at radii greater of $r \approx 14.1\arcmin$, so we extended the annuli to the complete FOV. This fraction is shown
in Figure~\ref{surfdensfield} ({\em middle panel}). The asymptotic value (at larger distances) of the logarithm of this fraction approaches
$-0.69 \pm 0.05$. Combining this number with the surface density of field stars outside the box, we obtain that the surface density of field
stars is $7 \, {\rm stars/arcmin^2}$, comprised of $5.8 \, {\rm stars/arcmin^2}$ outside the box and $1.2 \, {\rm stars/arcmin^2}$ inside the
box. The density profile of the cluster is shown in the bottom panel of Figure~\ref{surfdensfield}. The crosses correspond to the total density
profile of the cluster, and the filled circles are the surface density profile with the contamination of field stars subtracted.
The error bars correspond to the Poissonian error in the number counts. In fact, the profile does not seem to break at $r=14.1\arcmin$,
nor show any indication of flattening at large distances.

An analysis of the CMD at different radial distances is shown in Figure~\ref{radii}. Here, no selection criteria were used for the stars. 
The selected annuli correspond to different values of the tidal radius of M72 found in literature: $r_t \approx 7.4\arcmin$ is taken from 
the Dec. 2010 version of the Harris 1996 catalog; this value was calculated based on the listed value of the central concentration
$c=log(r_t/r_c) = 1.21$, with $r_c = 0.46\arcmin$.  The value $r_t \approx 9.2\arcmin$ is taken from Trager et al. 1995, and $14.1\arcmin$
is the limit of our study. Figure~\ref{radii} shows that there is still a trace of (mostly main sequence, MS) stars belonging to the cluster
at radial distances further than $r_t \approx 7.4\arcmin$ and even further than $r_t \approx 9.2\arcmin$.

The existence of an extratidal component in \objectname{NGC~6981} was previously claimed in Grillmair et al. (1995), along with several other
GGCs. They conclude that these stars are probably unbound due to ongoing stripping episodes; furthermore, they speculate that GCs
may not have an observable limiting radii. This is also observed in $N$-body simulations (e.g., Combes et al. 1999; Capuzzo Dolcetta et al.
2005; K\"upper et al. 2010, 2012). In our study, the derived density profile does not show flattening at larger distances. 
To put stronger constraints on the origin of the extratidal component of \objectname{NGC~6981}, covering more extended fields around 
the cluster would prove of great interest.

\section{Color-Magnitude Diagram}\label{sec:cmd}
In Figure~\ref{accbox} we present the results from our photometry in the $V$, $B-I$ plane for all stars ($\approx 33,\!000$)
detected in the three filters, throughout the field, with no selection criteria applied. The presence of field stars and background galaxies
is clear, especially in the zone occupied by red giant branch (RGB) stars and the redder and fainter part of the CMD. To produce a tighter CMD, 
we applied standard selection criteria used before in similar studies based on ALLFRAME photometry (see Stetson et al. 2003). 
The initial cut was to consider stars with $\sigma(B-I) \leq$ 0.1. Stars with values $\sigma(B-I)$ larger than 0.1~mag are mostly very faint
stars ($V \gtrsim 23$). We then used the ALLFRAME index \texttt{sharp}, which gives an estimate of the residuals from the PSF fit.
We selected stars with  $-1 \leq \texttt{sharp} \leq 1$, which removed the contamination due mostly to background galaxies
and very poorly measured stars. In order to obtain an accurate ridgeline and to study the more evolved populations, we applied one
more cut, based on the radial distance of the stars from the cluster center (e.g., Stetson et al. 2005). Figure~\ref{Vradius}
shows the distribution in magnitude of stars as a function of radial distance. The upper plot shows stars with $\sigma (B-I) \leq 0.1$, 
whereas the lower panel shows those with $0.1 \leq \sigma (B-I) \leq 0.3$. For the sake of clarity, we have marked in the upper plot
the limits at $V = 20$, $r =100\arcsec$ and $r=316\arcsec$. This figure suggests that the detection limit
for stars with good photometry within the magnitude range $18 \leq V \leq 20$ is at a radius of $\approx 30\arcsec$. Moreover, we can consider
that the detection limit for stars with $V \geq 20$ is constant from radii greater than $100\arcsec$ until $\approx 316\arcsec$, where 
there appears to be a break before field contamination increases at fainter magnitudes. We summarize all these considerations by applying
the following radial selection criteria:

\begin{eqnarray}
\begin{array}{rrl}
 V \leq 18  &  \,\,\, {\rm and} \,\,\,  & r \leq 316\arcsec, \\
18 \leq V \leq 20 & \,\,\, {\rm and} \,\,\, & 30\arcsec \leq r \leq 316\arcsec, \\
20 \leq V & \,\,\, {\rm and} \,\,\, & 100\arcsec \leq r \leq 316\arcsec. \\
\end{array}
\end{eqnarray}

\noindent We remark that, as stated in the previous section, the limit of $316\arcsec$ does not correspond to the cluster's tidal limit,
and it is used only for the purpose of obtaining a cleaner CMD.

In Figure~\ref{BVI}, we plot the cleaner CMD in all available filters, obtained after applying the described selection criteria.
The CMD reveals an overall morphology in good agreement with previous studies (Dickens 1972; Piotto et al. 2002; Bramich et al. 2011),
but this study reveals a more extended MS, reaching down to $V \approx 24$, and with errors less than $\sim 0.1$~mag at this faint level. 
Also, we note the presence of a well-defined RGB with moderate steepness, a horizontal branch (HB) with both red and blue populations,
a relatively populated asymptotic giant branch sequence, and the presence of a BSS population.

\begin{table}[t]
\begin{center}
\footnotesize
\caption{\footnotesize{Mean Fiducial Points for NGC~6981: $V$, $\bv$ plane.}}
\begin{tabular}{lc}
\tableline\tableline
\emph{V} & (\bv)  \\
\tableline
\tableline
\multicolumn{2}{c}{MS + RGB}\\
\tableline
$0.79$ & $23.10$ \\
$0.61$ & $22.01$ \\
$0.52$ & $21.36$ \\
$0.48$ & $20.74$ \\
$0.47$ & $20.30$ \\
$0.48$ & $20.01$ \\
$0.51$ & $19.85$ \\
$0.56$ & $19.74$ \\
$0.59$ & $19.70$ \\
$0.64$ & $19.60$ \\
$0.65$ & $19.54$ \\
$0.68$ & $19.30$ \\
$0.72$ & $18.70$ \\
$0.75$ & $18.00$ \\
$0.78$ & $17.40$ \\
$0.84$ & $16.80$ \\
$0.91$ & $16.20$ \\
$1.03$ & $15.50$ \\
$1.20$ & $14.70$ \\
$1.30$ & $14.30$ \\
$1.42$ & $14.00$ \\
 \tableline
\end{tabular}
\label{tab:fidpoints1}
\end{center}
\end{table}       

\begin{table}[t]
\begin{center}
\footnotesize
\caption{\footnotesize{Mean Fiducial Points for NGC~6981: $V$, $B-I$ plane.}}
\begin{tabular}{lc}
\tableline\tableline
\emph{V} & $(B-I)$  \\
\tableline
\tableline
\multicolumn{2}{c}{MS + RGB}\\
\tableline
$1.90$ &  $23.55$  \\
$1.78$ &  $23.20$  \\
$1.64$ &  $22.80$  \\
$1.55$ &  $22.53$  \\
$1.45$ &  $22.22$  \\
$1.36$ &  $21.92$  \\
$1.30$ &  $21.70$  \\
$1.23$ &  $21.41$  \\
$1.17$ &  $21.05$  \\
$1.13$ &  $20.80$  \\
$1.12$ &  $20.60$  \\
$1.11$ &  $20.40$  \\
$1.12$ &  $20.21$  \\
$1.14$ &  $20.03$  \\
$1.17$ &  $19.92$  \\
$1.23$ &  $19.80$  \\
$1.30$ &  $19.72$  \\
$1.37$ &  $19.66$  \\
$1.45$ &  $19.59$  \\
$1.50$ &  $19.44$  \\
$1.56$ &  $19.04$  \\
$1.62$ &  $18.57$  \\
$1.68$ &  $17.90$  \\
$1.75$ &  $17.37$  \\
$1.82$ &  $16.88$  \\
$1.91$ &  $16.46$  \\
$2.00$ &  $16.09$  \\
$2.11$ &  $15.68$  \\
$2.22$ &  $15.35$  \\
$2.36$ &  $14.98$  \\
$2.47$ &  $14.73$  \\
$2.60$ &  $14.50$  \\
$2.75$ &  $14.25$  \\
$2.87$ &  $14.13$  \\
 \tableline
\end{tabular}
\label{tab:fidpoints2}
\end{center}
\end{table}    

\begin{table}[t]
\begin{center}
\footnotesize
\caption{\footnotesize{Mean Fiducial Points for NGC~6981: $V$, $V-I$ plane.}}
\begin{tabular}{lc}
\tableline\tableline
\emph{V} & $(V-I)$  \\
\tableline
\tableline
\multicolumn{2}{c}{MS + RGB}\\
\tableline
$0.96$ &  $23.10$  \\
$0.86$ &  $22.51$  \\
$0.78$ &  $22.03$  \\
$0.70$ &  $21.30$  \\
$0.64$ &  $20.66$  \\
$0.64$ &  $20.28$  \\
$0.65$ &  $20.04$  \\
$0.68$ &  $19.86$  \\
$0.72$ &  $19.75$  \\
$0.76$ &  $19.70$  \\
$0.79$ &  $19.62$  \\
$0.83$ &  $19.52$  \\
$0.85$ &  $19.38$  \\
$0.87$ &  $19.05$  \\
$0.89$ &  $18.63$  \\
$0.91$ &  $18.12$  \\
$0.94$ &  $17.62$  \\
$0.99$ &  $16.90$  \\
$1.04$ &  $16.34$  \\
$1.11$ &  $15.73$  \\
$1.18$ &  $15.23$  \\
$1.26$ &  $14.82$  \\
$1.36$ &  $14.43$  \\
$1.42$ &  $14.21$  \\
$1.51$ &  $14.00$  \\

 \tableline
\end{tabular}
\label{tab:fidpoints3}
\end{center}
\end{table}    

As in many previous studies, including for instance Stetson et al. (2005) and Zorotovic et al. (2009),  
the mean ridgelines for the \bv, $V-I$, $B-I$ versus $V$ planes were determined by eye in large-scale plots. 
Then the normal points were overplotted in the CMD to perform some minor adjustments to obtain a smooth ridgeline. 
The fiducial points are presented in Tables~\ref{tab:fidpoints1} to \ref{tab:fidpoints3}, and the resulting 
rigeline, in the $B-I$ versus $V$ plane, is overplotted on the data in Figure~\ref{radii}.
We note that the apparent asymmetry of the MS around the fiducial line is probably due to binary sequence. Sollima et al. (2007), based
on HST/ACS photometry, estimated a binary fraction for \objectname{NGC~6981} of $\xi \sim 10\%$. 

The zero-age horizontal branch (ZAHB) level was determine in the same way as in Ferraro et al.\ (1999, hereafter F99; see their
Figure~3), by fitting the Victoria-Regina ZAHB models (VandenBerg et al.\ 2006) at $\log T_{\rm eff} = 3.85$, assuming $[\rm Fe/H] =  -1.41$
and $[\alpha/ \rm Fe] = +0.3$ (recall that the cluster has $[{\rm Fe/H}] = -1.42$, according to the December 2010 version of the Harris 1996
catalog, and that halo stars typically have a similar such level of $\alpha$-elements enhancement;  see, e.g., Pritzl et al.\ 2005).  
We matched the ZAHB in the dereddened CMD (using a value of $E($\bv$) = 0.05$; see Section~\ref{sec:metal} for details) by allowing vertical 
shifts. The result is shown in Figure~\ref{ZAHB}. In this way we found $V_{\rm ZAHB} = 16.89 \pm 0.02$~mag, in good agreement with the value
listed in F99 for M72. The turn-off point (TO) was determined by fitting a parabola to the data points in a small area of the MS around the TO.
We obtain $V_{\rm TO} = 20.31 \pm 0.1$~mag at a color $(B-I)_{\rm TO}= 0.47$~mag.

\subsection{Reddening and Metallicity}\label{sec:metal}
The Dec. 2010 version of the Harris (1996) catalog lists the reddening for \objectname{NGC~6981} to be $E(\bv) = 0.05$, 
based on previous photometric studies. For comparison, the Schlegel et al. (1998) dust maps give $E(\bv) = 0.06$. 
We adopted the Harris value for the remainder of our analysis.

F99 reviewed observational indicators that are derived from the evolved sequences in the CMD and can be used to 
estimate photometric metallicities (see their Table~4). We used the dereddened \bv, $V$ plane to measure the parameters
$(\bv)_{0,g}$ (intrinsic color of the RGB at the HB level), $S_{2.0}$, $S_{2.5}$ (intersection of the line connecting the point of intersection 
between the RGB and the HB level and the point 2.0 and 2.5 magnitudes brighter than the HB), $\Delta V_{1.1}$, and $\Delta V_{1.2}$ (height of
the RGB brighter than the HB at color $(\bv)_0 = 1.1$, 1.2), as shown in Figure~\ref{ZAHB}.
The very red star close to the RGB tip level, at a color $\bv \approx 1.7$, 
corresponds to variable star V42 (Figure~\ref{CMDvar}). This star is found to vary, but the
type of variability is not identified and it may not be a member of the
cluster (see sec.~\ref{sec:variables} for details). For this reason, we avoid extrapolating the ridgeline
to reach this star, and the parameter $\Delta V_{1.4}$ was not used in this work. 

The values of the parameters obtained in our study as well as the resulting values for $\rm [Fe/H]$ 
(in the Carretta \& Gratton 1997 [CG97] metallicity scale) and $\rm [M/H]$ are summarized in Table~\ref{tab:met}. Taking the mean, the final
adopted values are  $\rm [M/H] = -1.19 \pm 0.05$ and  $\rm [Fe/H] = -1.32 \pm 0.17$.

It should be noted that recently, Carretta et al. (2009) presented a recalibration of the CG97 metallicity scale,
based on a homogeneous analysis of a large sample of spectra taken with UVES. The relation between both scales is 

\begin{equation}
 \rm [Fe/H]_{UVES} = 1.137 (\pm 0.060)\rm [Fe/H]_{CG97} - 0.003, 
\end{equation}

\noindent as provided in their study. 
The CG97 values were transformed to this scale, and are also summarized in Table~\ref{tab:met}. The final metallicity value 
adopted in this paper, in the UVES scale is $\rm [Fe/H] = -1.50 \pm 0.19$. 
Similarly, we transformed the values of $\rm [M/H]$ in Table~\ref{tab:met} to the UVES metallicity scale by using the theorethical equation
found by Salaris, Chieffi \& Straniero (1993), used in F99, this time using the new UVES $\rm [Fe/H]$ values

\begin{equation}
 {\rm [M/H]_{UVES}} = {\rm [Fe/H]_{UVES}} + \log (0.638 \, f_{\alpha} + 0.362),
\end{equation}

\noindent where $f_{\alpha}$ is the enhancement factor of the $\alpha$-elements; in this case we use $f_{\alpha} = 10^{0.28}$, as suggested in
F99. The results are shown in column 5 of Table~\ref{tab:met}.

Since the slopes $S_{2.0}$ and $S_{2.5}$ do not depend on the adopted reddening, we use those values and the obtained value of $\rm [Fe/H]$ 
to find, for the reddened color of the RGB at the HB level, a value $(\bv)_{g} = 0.82$. This leads to a reddening value of $E(\bv) = 0.039$, 
in good agreement with our adopted value. 

 Other methods to determine the reddening are based in the minimum light colors of ab-type RR Lyrae stars (description of the RR Lyrae population
and other variables is in Section~\ref{sec:variables}). Indeed, over the phase  range $0.5 < \phi < 0.8$, the intrinsic colors of RRab's 
are fairly uniform (e.g., Preston \& Spinrad 1959; Preston 1964). Therefore, we   also estimated the reddening using Sturch's method 
(Sturch 1966; Walker 1990), a relation between color at minimum light, period, and  (more weakly) metallicity, as described by Walker (1998), 
where the reddening is calculated with the following relation:
 
 \begin{equation}
  E(\bv) = (\bv)_{\rm min} -0.24 \, P - \rm 0.056[Fe/H] -0.036,
 \end{equation}
 
 \noindent where $(\bv)_{\rm min}$ is the color of the RRab star at minimum light, $P$ is the fundamental period
 of the variable in days, and $\rm [Fe/H]$ is the metallicity in the Zinn \& West (1984; ZW84) scale. We use our result listed in 
Table~\ref{tab:met} in the UVES  scale, and transformed to the ZW84 scale using the second-order polynomial relation provided by 
Carreta et al. (2009), giving a value of $\rm [Fe/H] = -1.57$.
 By only considering stars with $\sigma(\bv)_{\rm min} < 0.06$~mag, we obtain a reddening value of $E(\bv) = 0.08$~mag, which is
 slightly higher than our  adopted value. 
 As shown by Guldenschuh et al. (2005), the $(V-I)_{\rm 0, min}$ colors of ab-type RR Lyrae stars are also remarkably uniform, with  
 $(V-I)_{\rm 0,min} = 0.58 \pm 0.02$~mag, irrespective of period, amplitude, and metallicity (Kunder et al. 2013). We calculated the 
 $V-I$ color at minimum light from our data, and on this basis we found a mean value of $E(V-I) = 0.07$~mag, or $E(\bv) = 0.06$~mag, 
 assuming $E(V-I) = 1.27 \, E(\bv)$~-- as obtained from eq.~1 in Dean et al. (1978), for $(\bv)_0 \approx 0.3$~mag (as appropriate 
 for RR Lyrae stars) and $E(\bv) \approx 0.05$~mag. This result is in excellent agreement with our adopted value. 
 It is worth noting that higher reddening values for \objectname{M72} have on occasion also favored in the literature, 
 including for instance $E(\bv) = 0.07$~mag (de Santis \& Cassisi 1999) and  $E(\bv)= 0.11$~mag (Rodgers \& Harding 1990).

\begin{table*}[t]
\begin{center}
\footnotesize
\caption{\footnotesize{RGB parameters for NGC~6981}}
\begin{tabular}{lcccc}
\tableline\tableline
Parameter               & ${\rm [Fe/H]_{CG97}}$ & ${\rm [M/H]}_{\rm CG97}$  & ${\rm [Fe/H]_{UVES}}$ & ${\rm [M/H]}_{\rm UVES}$ \\
\tableline
$(\bv)_{0,g} = 0.781$   & $-1.42$  & $-1.21$     & $-1.61$ & $-1.42$\\
$\Delta V_{1.1} = 1.96$	& $-1.29$  &	$-1.10$  & $-1.47$ & $-1.28$\\
$\Delta V_{1.2} = 2.39$	& $-1.38$  &	$-1.18$  & $-1.57$ & $-1.38$\\
$S_{2.0} = 6.07$        & $-1.03$  &	$-1.23$  & $-1.18$ & $-0.98$\\
$S_{2.5} = 5.59$        & $-1.48$  &	$-1.23$  & $-1.69$ & $-1.49$\\
Mean                    & $-1.32 $ & $-1.19 $    & $-1.50$ & $-1.31$\\
\tableline 
\end{tabular}
\label{tab:met}
\end{center}
\end{table*} 

\begin{center}
 \begin{deluxetable*}{lcccccccccl}
 \tablewidth{0pc} 
 \tabletypesize{\scriptsize}
 \tablecaption{Total census of variable star population in NGC~6981}
 \tablehead{\colhead{ID} & \colhead{RA (J2000)} & \colhead{DEC (J2000)} & \colhead{$P$} & \colhead{$A_B$} & \colhead{$A_V$} & \colhead{$A_I$} & \colhead{$\langle B \rangle$} & \colhead{$\langle V \rangle$} & \colhead{$\langle I \rangle$} & \colhead{Comments} \\
 \colhead{} & \colhead{(h:m:s)} & \colhead{(deg:m:s)} & \colhead{(days)} & \colhead{(mag)} & \colhead{(mag)} & \colhead{(mag)} & \colhead{(mag)} & \colhead{(mag)} & \colhead{(mag)} & \colhead{} }
 \startdata
 V1	& $20:53:31.11$& $-12:33:11.6$ & $0.619784$ & $0.81$ & $0.62$ & $0.37$ & $17.28$ & $16.87$ & $16.28$  &\RRab\ \\
 V2	& $20:53:34.56$& $-12:29:02.0$ & $0.465256$ & $1.58$ & $1.20$ & $0.75$ & $17.18$ & $16.87$ & $16.45$  & \RRab\ \\
 V3	& $20:53:24.57$& $-12:33:17.0$ & $0.497602$ & $1.18$ & $0.88$ & $0.58$ & $17.11$ & $16.79$ & $16.30$  & \RRab\ \\
 V4	& $20:53:20.80$& $-12:31:42.4$ & $0.552487$ & $1.14$ & $0.91$ & $0.59$ & $17.27$ & $16.89$ & $16.33$  &\RRab\ \\
 V5	& $20:53:25.58$& $-12:32:41.4$ & $0.507262$ & $1.32$ & $0.96$ & $0.59$ & $17.09$ & $16.74$ & $16.26$  &\RRab\  \\
 V7	& $20:53:27.77$& $-12:31:20.9$ & $0.524683$ & $1.16$ & $0.87$ & $0.55$ & $17.19$ & $16.84$ & $16.35$ &\RRab\ \\
 V8	& $20:53:27.58$& $-12:30:47.7$ & $0.568377$ & $1.05$ & $0.81$ & $0.54$ & $17.27$ & $16.88$ & $16.34$ &\RRab\ \\
 V9	& $20:53:28.66$& $-12:31:27.8$ & $0.602929$ & $0.88$ & $0.64$ & $0.40$ & $17.29$ & $16.88$ & $16.30$ &\RRab\ \\ 
 V10	& $20:53:24.85$& $-12:33:30.8$ & $0.558182$ & $1.10$ & $0.90$ & $0.58$ & $17.25$ & $16.89$ & $16.34$  &\RRab\ \\
 V11	& $20:53:32.05$& $-12:32:51.1$ & $0.520638$ & $1.12$ & $0.82$ & $0.46$ & $17.23$ & $16.87$ & $16.35$  &\RRab, Bl \\
 V12	& $20:53:28.60$& $-12:32:38.4$ & $0.287861$ & $0.61$ & $0.47$ & $0.28$ & $17.00$ & $16.72$ & $16.32$  &\RRc\ \\
 V13	& $20:53:28.91$& $-12:32:01.9$ & $0.542034$ & $1.09$ & $0.86$ & $0.53$ & $16.94$ & $16.56$ & $16.03$  &\RRab\ \\
 V14	& $20:53:27.17$& $-12:31:42.6$ & $0.607213$ & $0.86$ & $0.65$ & $0.39$ & $17.17$ & $16.76$ & $16.18$  &\RRab, Bl \\
 V15	& $20:53:23.76$& $-12:32:38.6$ & $0.540480$ & $1.19$ & $0.85$ & $0.56$ & $17.21$ & $16.87$ & $16.33$  &\RRab, Bl \\
 V16	& $20:53:27.86$& $-12:32:36.6$ & $0.575212$ & $1.01$ & $0.77$ & $0.49$ & $17.18$ & $16.79$ & $16.23$  &\RRab\ \\
 V17	& $20:53:28.23$& $-12:32:59.7$ & $0.573540$ & $1.07$ & $0.84$ & $0.53$ & $17.24$ & $16.86$ & $16.32$  &\RRab\ \\
 V18	& $20:53:26.22$& $-12:32:54.4$ & $0.535576$ & $1.01$ & $0.83$ & $0.56$ & $16.91$ & $16.62$ & $16.18$  &\RRab\ \\
 V20	& $20:53:24.22$& $-12:32:03.7$ & $0.595047$ & $0.95$ & $0.71$ & $0.44$ & $17.28$ & $16.88$ & $16.29$ &\RRab\ \\
 V21	& $20:53:22.40$& $-12:32:05.5$ & $0.531161$ & $1.28$ & $1.07$ & $0.70$ & $17.27$ & $16.92$ & $16.39$ &\RRab\ \\
 V23	& $20:53:21.15$& $-12:30:21.9$ & $0.585121$ & $0.85$ & $0.67$ & $0.46$ & $17.30$ & $16.91$ & $16.33$  &\RRab, Bl \\
 V24	& $20:53:27.18$& $-12:32:41.9$ & $0.327127$ & $0.43$ & $0.29$ & $0.14$ & $16.83$ & $16.42$ & $15.87$  &\RRc\ \\
 V25	& $20:53:18.90$& $-12:31:13.6$ & $0.353350$ & $0.54$ & $0.42$ & $0.27$ & $17.11$ & $16.80$ & $16.35$  &\RRc\ \\
 V27	& $20:53:42.60$& $-12:36:06.4$ & $0.673871$ & $1.37$ & $1.11$ & $0.66$ & $16.95$ & $16.62$ & $16.10$ &\RRab\ \\
 V28	& $20:53:32.26$& $-12:30:55.7$ & $0.567251$ & $0.82$ & $0.69$ & $0.48$ & $17.27$ & $16.88$ & $16.34$ &\RRab, Bl \\
 V29	& $20:53:25.77$& $-12:33:11.2$ & $0.597448$ & $1.04$ & $0.87$ & $0.54$ & $17.23$ & $16.85$ & $16.29$ &\RRab\ \\
 V31	& $20:53:28.20$& $-12:31:42.7$ & $0.542349$ & $0.74$ & $0.69$ & $0.51$ & $17.20$ & $16.80$ & $16.27$   &\RRab, Bl \\
 V32	& $20:53:18.84$& $-12:33:01.2$ & $0.528315$ & $1.26$ & $0.98$ & $0.62$ & $17.25$ & $16.91$ & $16.39$   &\RRab, Bl \\
 V35	& $20:53:43.56$& $-12:31:52.1$ & $0.543749$ & $1.05$ & $0.84$ & $0.52$ & $17.25$ & $16.89$ & $16.37$ &\RRab, Bl \\
 V36	& $20:53:26.90$& $-12:32:17.6$ & $0.582612$ & $1.02$ & $0.70$ & $0.46$ & $17.09$ & $16.70$ & $16.12$ &\RRab, Bl \\
 V39	& $20:53:41.06$& $-12:28:15.7$ & $0.426785$ & $0.68$ & $0.46$ & $0.29$ & $17.73$ & $17.30$ & $16.70$ &\RRab, f?, Bl \\
 V42	& $20:53:28.86$& $-12:32:16.5$ & $1.01109 $ & $0.47$ & $0.36$ & $0.17$ & $15.53$ & $13.85$ & $12.14$ &?? \\
 V43	& $20:53:27.35$& $-12:32:22.1$ & $0.283497$ & $0.56$ & $0.49$ & $0.30$ & $16.97$ & $16.74$ & $16.42$ &\RRc\ \\
 V44	& $20:53:28.02$& $-12:32:29.4$ & $0.557440$ & $1.16$ & $0.89$ & $0.60$ & $17.25$ & $16.87$ & $16.32$ &\RRab\ \\
 V45	& $20:53:28.66$& $-12:32:20.0$ & $0.364981$ & $0.64$ & $0.49$ & $0.28$ & $17.04$ & $16.77$ & $16.38$ &\RRc\ \\
 V46	& $20:53:28.96$& $-12:32:26.2$ & $0.286682$ & $0.37$ & $0.26$ & $0.15$ & $16.95$ & $16.67$ & $16.27$ &\RRc\ \\
 V47	& $20:53:29.73$& $-12:32:26.0$ & $0.649075$ & $0.32$ & $0.25$ & $0.18$ & $17.30$ & $16.86$ & $16.24$ &\RRab\ \\
 V48	& $20:53:26.46$& $-12:32:27.0$ & $0.639765$ & $0.59$ & $0.46$ & $0.28$ & $17.27$ & $16.84$ & $16.24$ &\RRab\ \\
 V49	& $20:53:28.27$& $-12:32:10.7$ & $0.578272$ & $1.00$ & $0.79$ & $0.50$ & $17.14$ & $16.73$ & $16.15$ &\RRab, Bl \\
 V50	& $20:53:28.25$& $-12:31:58.2$ & $0.488881$ & $1.29$ & $0.93$ & $0.61$ & $17.00$ & $16.61$ & $16.04$ &\RRab\ \\
 V51	& $20:53:28.42$& $-12:32:31.9$ & $0.548599$ & $0.90$ & $0.63$ & $0.31$ & $17.04$ & $16.58$ & $15.93$ &\RRab, Bl \\
 V52	& $20:53:27.97$& $-12:32:02.0$ & $0.698690$ & $0.74$ & $0.61$ & $0.36$ & $17.17$ & $16.72$ & $16.09$ &\RRab\ \\
 V53	& $20:53:27.00$& $-12:32:16.3$ & $0.652118$ & $0.57$ & $0.45$ & $0.25$ & $17.31$ & $16.87$ & $16.23$ &\RRab, Bl \\
 V54	& $20:53:28.64$& $-12:32:02.1$ & $0.077343$ & $0.38$ & $0.22$ & $0.16$ & $18.33$ & $17.95$ & $17.49$ & SX Phe \\
 V55	& $20:53:24.46$& $-12:31:27.1$ & $0.047033$ & $0.22$ & $0.17$ & $0.09$ & $19.65$ & $19.34$ & $18.95$ & SX Phe\\
 V56   & $20:53:28.92$& $-12:33:05.8$ & $0.048154$ & $....$ & $....$ & $....$ & $.....$ & $.....$ & $.....$ & SX Phe \\
 V57	& $20:53:27.36$& $-12:32:13.1$ & $0.335054$ & $0.44$ & $0.32$ & $0.17$ & $17.02$ & $16.63$ & $16.06$  &\RRc\ \\
 V59	& $20:53:48.89$& $-12:36:44.8$ & $0.603277$ & $0.91$ & $0.66$ & $0.44$ & $17.24$ & $16.86$ & $16.29$  &\RRab\ \\
 V60	& $20:53:46.68$& $-12:27:33.0$ & $0.482069$ & $1.73$ & $1.25$ & $0.77$ & $17.23$ & $16.94$ & $16.48$  &\RRab\ \\
 \enddata                                                                      
 \label{tab:variables}                                                         
 \end{deluxetable*}
 \end{center}

\subsection{RGB Bump}\label{sec:RGB}
An important feature of the RGB is the presence of a well-defined peak in its luminosity function (LF). This peak is known as the RGB
bump and was first predicted by Thomas (1967) and Iben (1968) as a consequence of the encounter of the H-burning shell and the chemical 
composition discontinuity left behind after the convective envelope reaches the maximum depth during the first dredge-up. This produces an
increase in the H-abundance, which causes a sudden drop in the mean molecular weight $\mu$. Since the efficiency of the H-burning
 shell is proportional to a high power of $\mu$, the luminosity decreases (Cassisi et al. 2011). Evolution of stars in this stage slows down 
while they try to adjust to the new conditions, resulting in an overdensity in the LF. The RGB bump, first detected by King et al. (1985),
is known to be located at higher luminosities for more metal-poor GCs (e.g., Fusi Pecci et al. 1990; Recio-Blanco \& Laverny 2007). 
The position of the RGB bump is a key quantity to study the evolutionary properties of RGB stars (e.g., Catelan 2007, and references therein.

To determine the position of the bump, we followed a method similar to the one described in Zoccali et al. (2001).
We select the RGB stars in the CMD, restricting ourselves to those stars within $r \leq 316\arcsec$, to avoid field 
contamination (Figure~\ref{historgb} ({\em top}). Although it is possible that some field stars are present in the selected RGB sample,
these should be sufficiently few that our final result is not affected. 

The luminosity function constructed for the selected stars is shown in Figure~\ref{historgb} ({\em middle}). We constructed an average-shifted
 histogram (ASH, Scott 1985) by combining 8 different histograms, each calculated in an interval in magnitude of $\Delta V = 4$ and with
a bin width of 0.25 but with different origin values $x_0 = 14, 14.2, 14.4, 14.6, 14.8$. This method of smoothing a dataset allows us to
 determine a significant peak in the data which does not depend on the choice of bin width and origin.  Figure~\ref{historgb} ({\em bottom})
 displays the cumulative distribution. The change in the slope corresponds to the position of the RGB bump. Based on these plots, 
the location of the bump was measured at $V_{\rm bump} = 16.6$.

F99 studied the relation between $\Delta V^{\rm bump}_{\rm HB}$ (difference in magnitude between the RGB bump and the ZAHB level) 
and the metallicity of the cluster, confirming a strong correlation between them. The relation is shown in Figure~\ref{bumpFeH}, for 
$\rm [Fe/H]$ and global metallicity $\rm [M/H]$ in the CG97 scale. The data were taken from Table~5 in F99. The solid circle
represents the value for \objectname{NGC~6981} found in this study, with $\Delta V^{\rm bump}_{\rm HB} =  -0.29$. Our result is 
clearly consistent with the correlation found in F99.

Another useful parameter involving the RGB bump is $R_{\rm bump}$, defined in Bono et al. (2001). 
This is the ratio between the number of RGB stars in the bump region (i.e., at $V_{\rm bump} \pm 0.4$) and the number of RGB stars 
in the interval  $V_{\rm bump} + 0.5 < V < V_{\rm bump} + 1.5$. 
This parameter is an indicator of the ocurrence of deep mixing before the 
H-shell encounters the chemical discontinuity left over by the first dredge-up, which would have an effect on the time spent in the bump region.
In Figure~\ref{Rbump} we plot the data taken from Bono et al. for $R_{\rm bump}$ as a function of global metallicity.
For \objectname{NGC~6981}, we obtained a value of $R_{\rm bump} = 0.61 \pm 0.10$ (solid circle in Figure~\ref{Rbump}). Bono et al. found
a large value of  $R_{\rm bump}$ for \objectname{NGC~6981}, mostly due to a small number of bump stars ($N_{\rm bump}= 40$, see their Table~1). 
We found instead $N_{\rm bump}= 58$, which leads to a slightly smaller $R_{\rm bump}$ value, more consistent~-- but still not completely so~-- 
with other clusters with similar metallicity. As found by Bono et al., this plot shows that the bulk of metal-poor clusters do not experience 
deep mixing before stars reach the bump.

\subsection{HB Morphology}\label{sec:HB}
The HB of NGC~6981 (Figure~\ref{ZAHB}) does not show any prominent feature, such as an extended HB or gaps in the blue part, so it can be 
well described by the HB index $\mathcal{L} = \mathcal{(B - R)/(B + V + R)}$ (Lee et al. 1994), were $\mathcal{B}$, $\mathcal{R}$, and 
$\mathcal{V}$ correspond to the numbers of blue, red, and variable (RR Lyrae-type) 
stars on the HB. To obtain the counts in each case, we selected stars within a radius of
100\arcsec, to avoid the contamination of field stars, particularly in the red HB; however, the inclusion of a couple of stars from the RGB,
scattered by photometric errors, might be possible. This yields a value of $\mathcal{L} = +0.02$, slightly less than the value 
of $\mathcal{L} = +0.14$  listed by Mackey \& van den Bergh (2005).  
Another significant parameter is $\mathcal{P}_{\rm HB} = \mathcal{(B{\rm 2} - R)/(B + V + R)}$  (Buonanno 1993), where $\mathcal{B{\rm 2}}$ 
corresponds to the number of stars in the HB with dereddened colors in the interval $-0.02 < (\bv)_0 < 0.18$. This parameter
was introduced in order to overcome the saturation of the $\mathcal{L}$ parameter for extreme blue HB morphologies. With the same constraints
as before, we found $\mathcal{P}_{\rm HB} = -0.27$ for M72.

\subsection{BSS Distribution}\label{sec:BSS}
BSS are a subpopulation present mostly in the cores of GCs. They appear as a brighter and hotter extension of the MS in the CMD, 
beyond the cluster's TO point, suggesting the presence of stars more massive than normal MS stars. 
There are currently two main scenarios for their origin: mass transfer 
between primordial binaries and direct collisions of stars in dense environments (see Ferraro \& Lanzoni 2009 for a recent review). These two
scenarios are non-exclusive, and can even coexist (Ferraro et al. 2009).  

Because of their high masses compared to normal MS stars, BSS are expected to be highly concentrated in the cluster centers. 
In fact, the radial distribution of BSS is typically bimodal, which is explained in terms of mass segregation and the nature 
of their progenitors (Mapelli et al. 2006; recent examples are found in Carraro \& Seleznev 2011 and Salinas et al. 2012), 
but in some cases, the radial distribution of BSS might be flat (see Beccari et al. 2011 and Contreras Ramos et al. 2012 for 
examples), indicating the presence of younger, not-yet-segregated populations.

The BSS population of \objectname{NGC~6981} is clearly identified in Figure~\ref{BSScmd}, framed by the selection box applied. We also selected 
a subsample from the RGB in order to compare both populations. The boxes were chosen in order to avoid the regions of the CMD with spurious
blends. For the determination of the cumulative radial distribution of BSS, we chose as a limit the tidal radius at $r_t =7.4\arcmin$, 
from Harris (2010; see sec.\ref{sec:extratidal}). The extratidal component found in Section~\ref{sec:extratidal} contains mostly MS stars
 and it is too small to study the BSS. Within the tidal radius, we found 39 BSS and 330 RGB stars. The cumulative radial distribution 
for both populations is displayed in Figure~\ref{BSScum}. This provides direct evidence of the more centrally concentrated distribution 
of the BSS, supporting mass segregation. A Kolmogorov-Smirnov test reveals that the probability of the two populations being drawn 
from the same parent distribution is less than 0.1\%.

\section{Relative Age Determination: Comparison with M3}\label{sec:age}
To obtain a relative age for \objectname{NGC~6981}, 
we compared this study with unpublished photometry for M3. This cluster has a similar 
metallicity (according to the Dec. 2010 version of the Harris 1996 catalog, $\rm [Fe/H]_{\rm M3} = -1.50$) and its CMD morphology resembles 
that of \objectname{NGC~6981}. The ridgeline and the TO point of M3 were determined using the same method described in Section~\ref{sec:cmd}
for \objectname{NGC~6981}. The $V$ magnitude of the HB was obtained using the Victoria-Regina ZAHB models, in the same way as with M72,  
assuming $[\rm Fe/H] =  -1.41$ and $[\alpha/ \rm Fe] = +0.3$. For M3, we determined the TO magnitude level and the HB level to be
at $V_{\rm TO} = 19.08 \pm 0.1$~mag and $V_{\rm HB} =  15.66 \pm 0.02$~mag, respectively. 
In the case of \objectname{NGC~6981}, we recall from Section~\ref{sec:cmd} the values $V_{\rm TO} = 20.31 \pm 0.1$~mag and 
$V_{\rm ZAHB} = 16.89 \pm 0.02$~mag.

  \begin{figure}[t]
   \centering
    \includegraphics[width=0.5\textwidth]{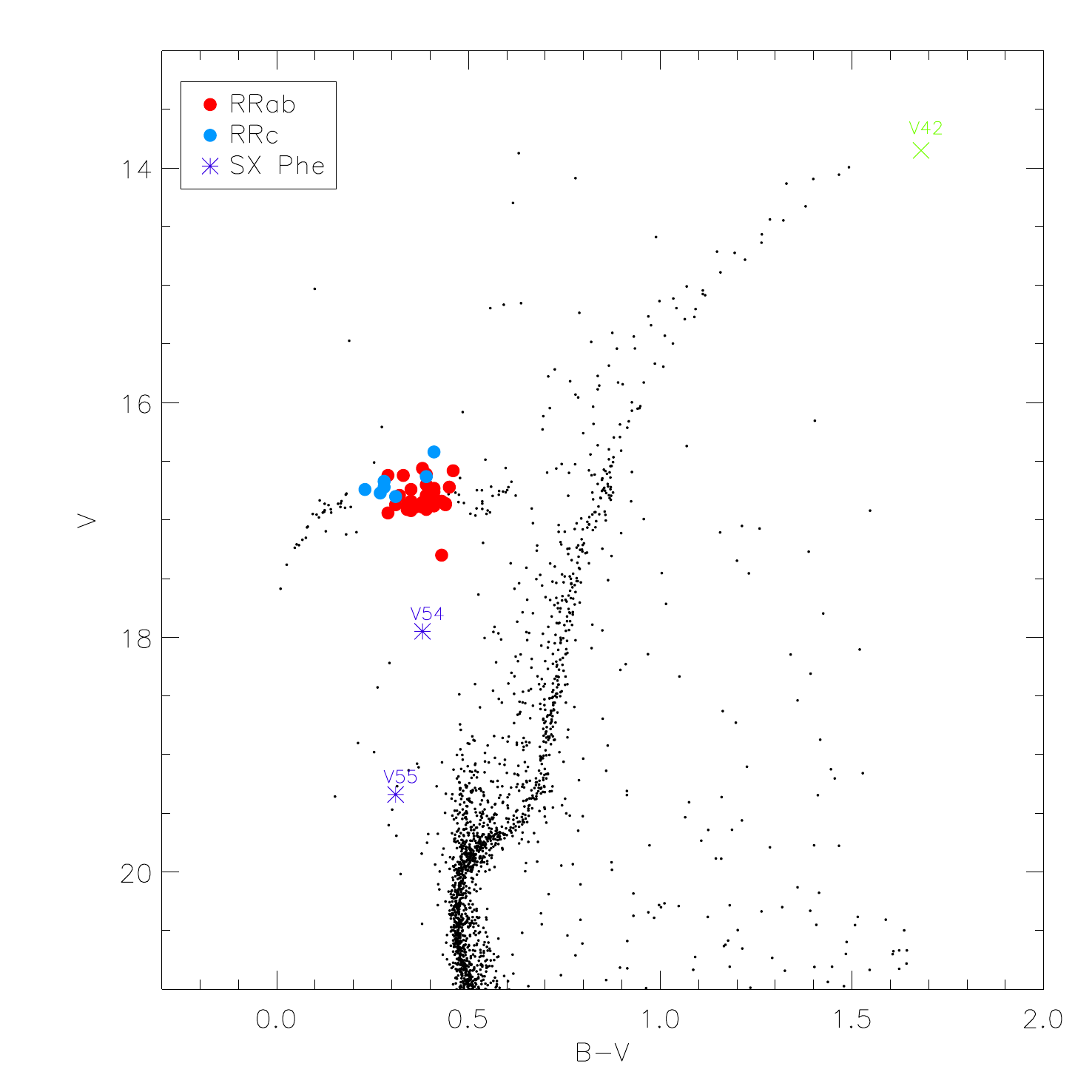}
   \caption{Positions of the detected variables in the CMD of NGC~6981. 
   RRab stars are shown as red circles, c-type RR Lyrae as blue circles. V54 and V55, two SX Phe stars, are shown as plus signs. 
   The green plus sign close to the RGB tip corresponds to the candidate variable V42 (see Sect.~\ref{sec:commentvar}
   for further details).}
   \label{CMDvar}
   \end{figure}

%
%
%
 %
   \begin{figure}[ht]
    \begin{center}
   \includegraphics[width=.45\textwidth]{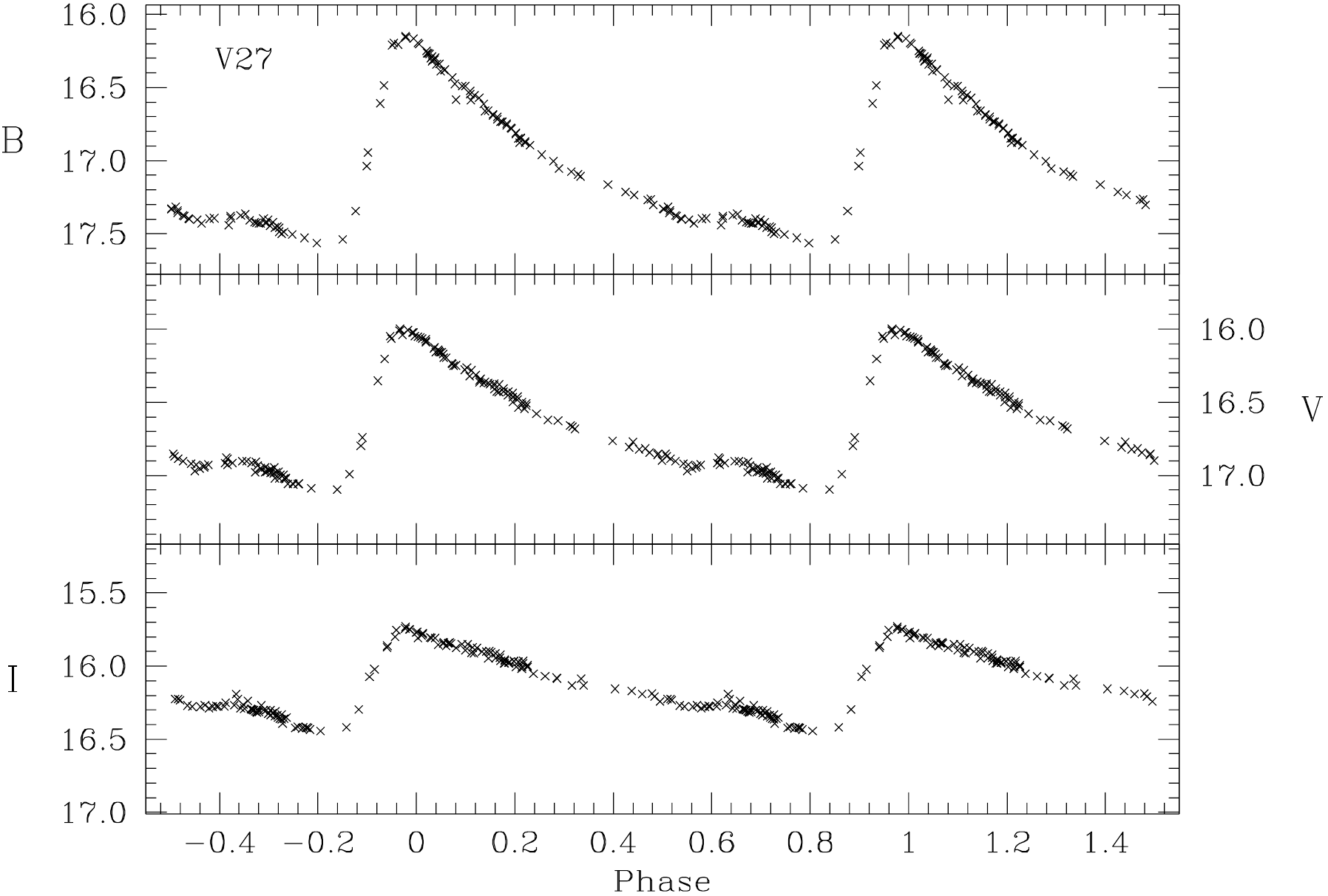}
   \includegraphics[width=.45\textwidth]{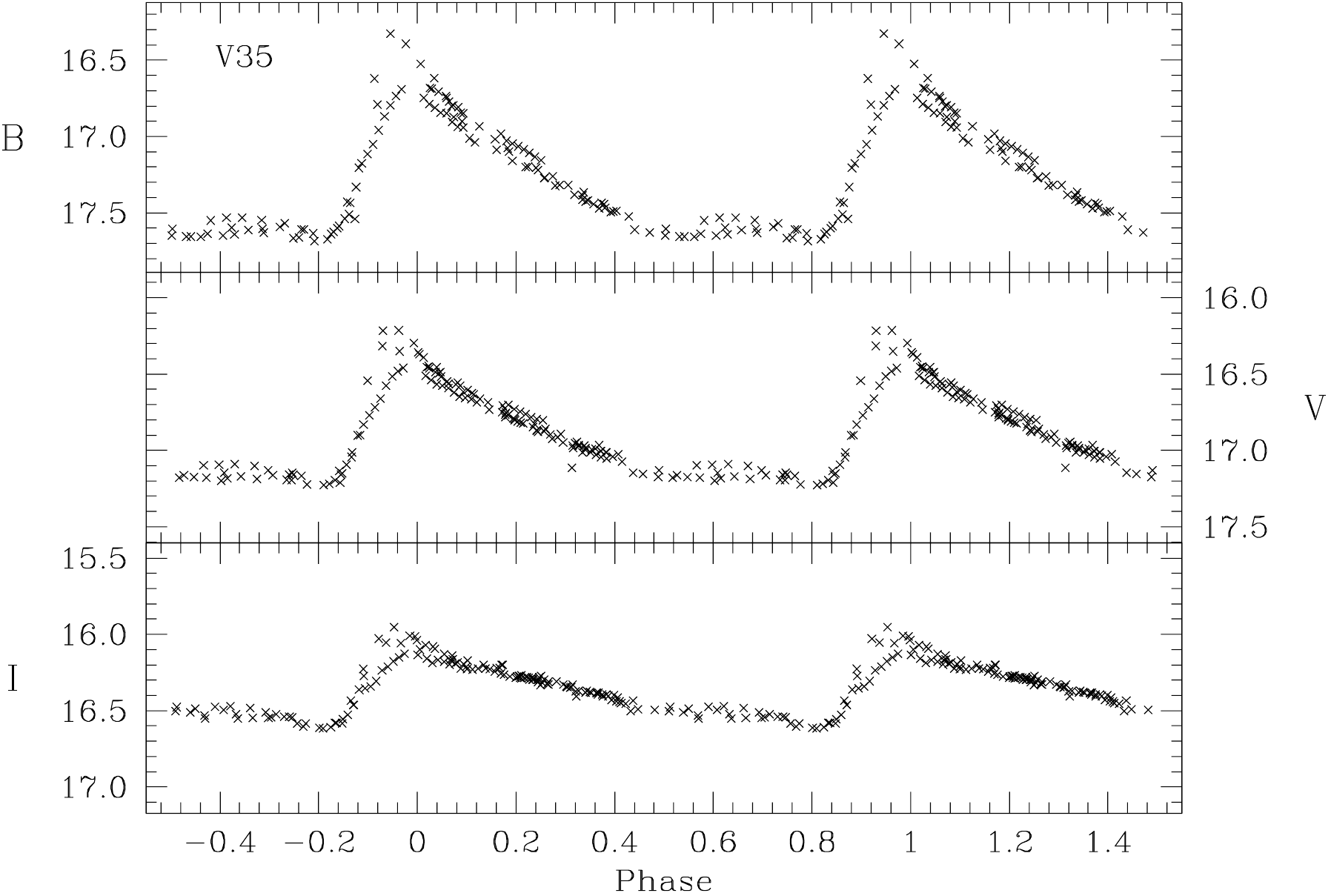}
    \includegraphics[width=.45\textwidth]{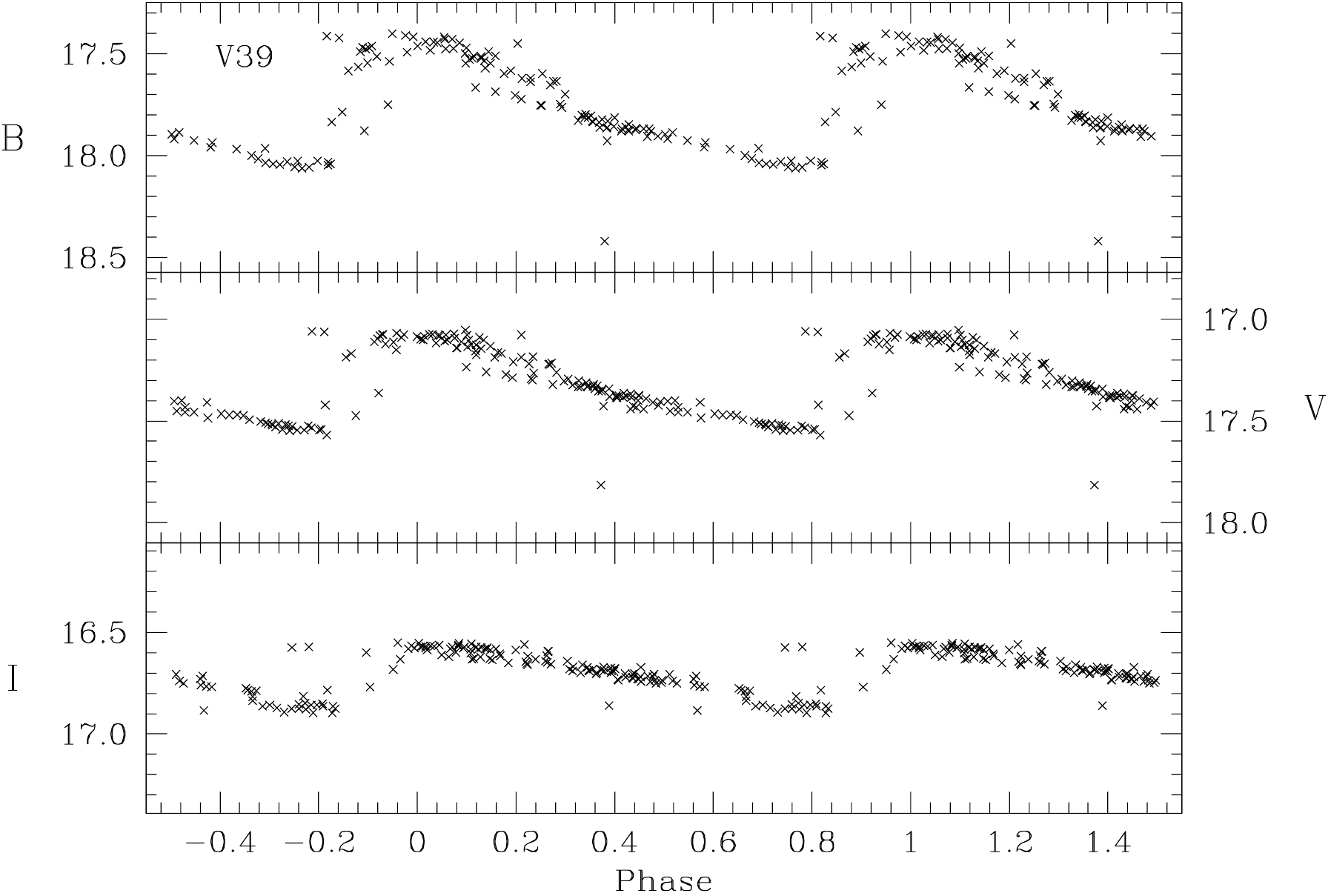}
    \includegraphics[width=.45\textwidth]{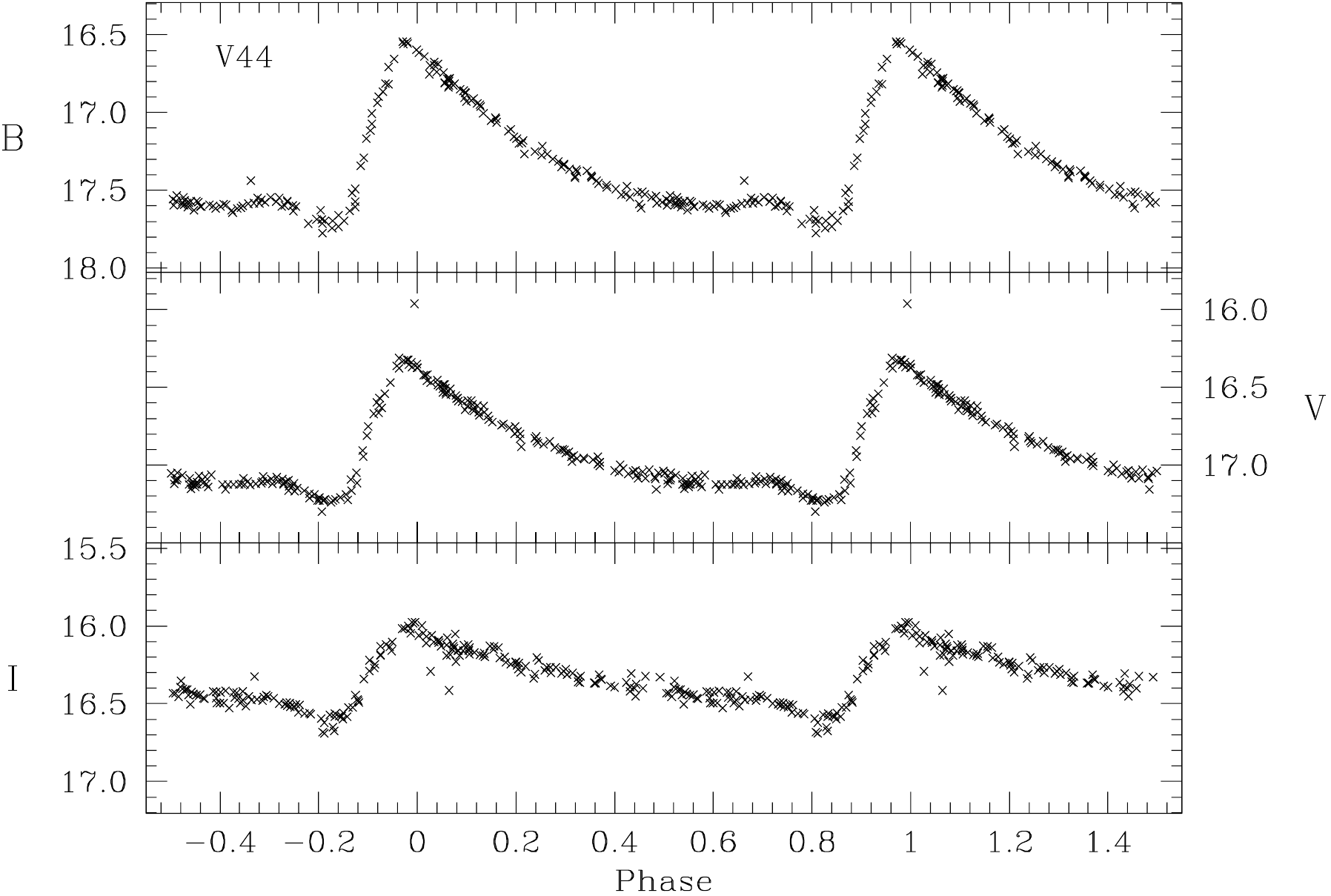}
     \caption{From top to bottom the light curves of variable stars 
     V27, V35, V39, and V44 are shown. For each individual star, from 
     top to bottom, the $B$, $V$, and $I$ light curves are displayed.}
    \label{figvar1}
    \end{center}
   \end{figure}
  %
   
    \begin{figure}
    \centering
     \includegraphics[width=.45\textwidth]{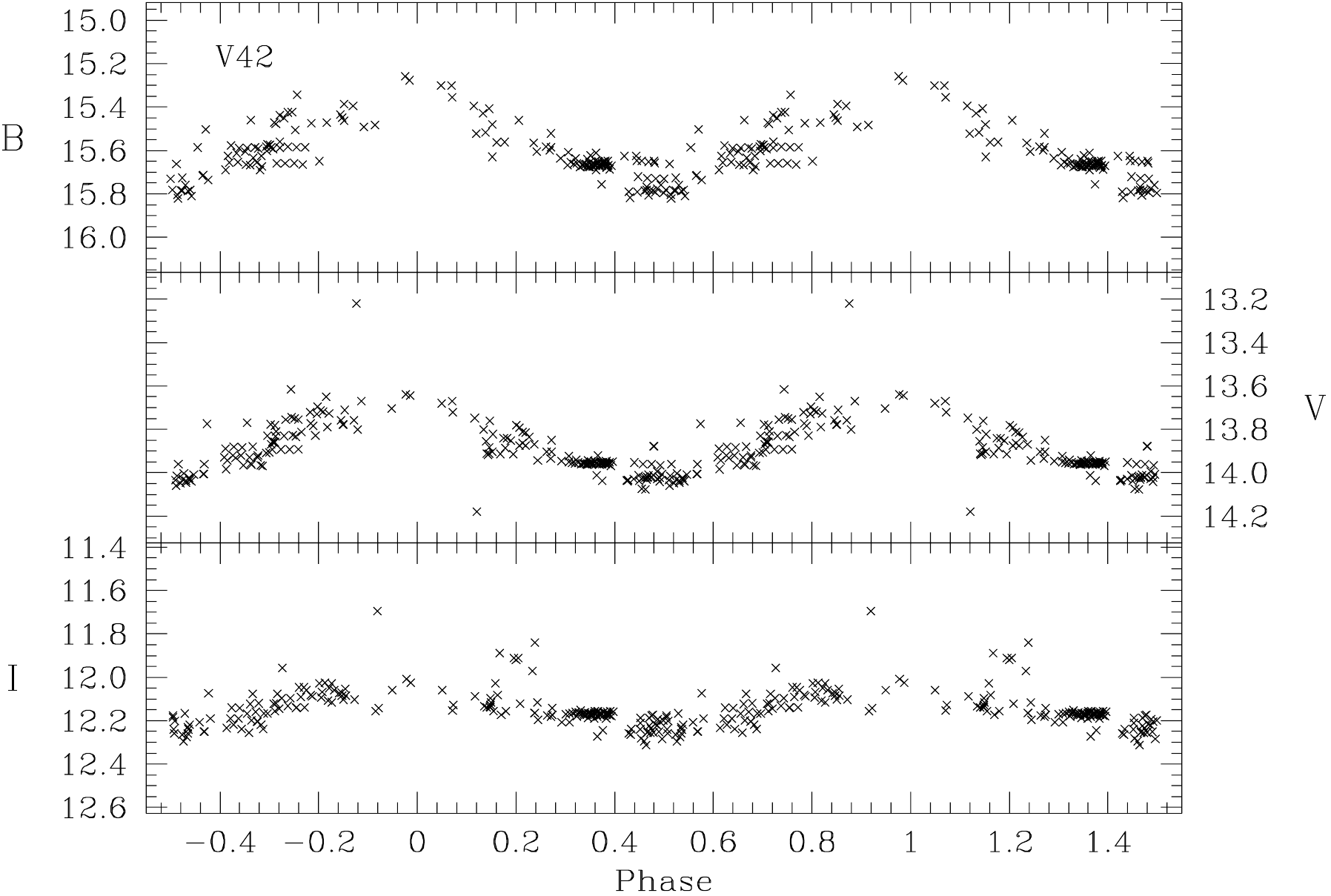}
    \caption{As in Figure~\ref{figvar1}, but for V42. 
     The period used is 1.011089~d.}
    \label{V42}
    \end{figure}
   
  %
    \begin{figure}
   \includegraphics[width=.45\textwidth]{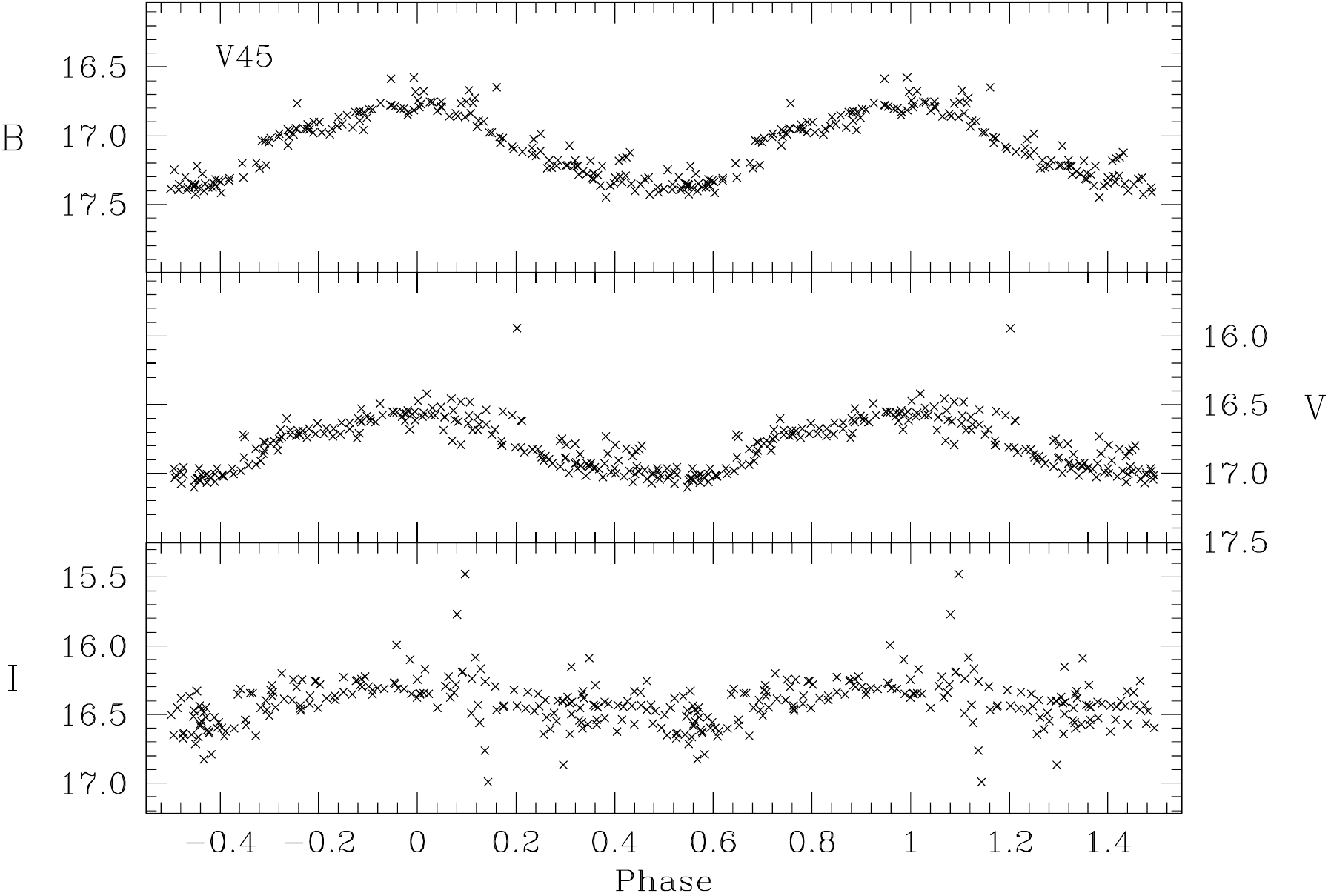}
     \includegraphics[width=.45\textwidth]{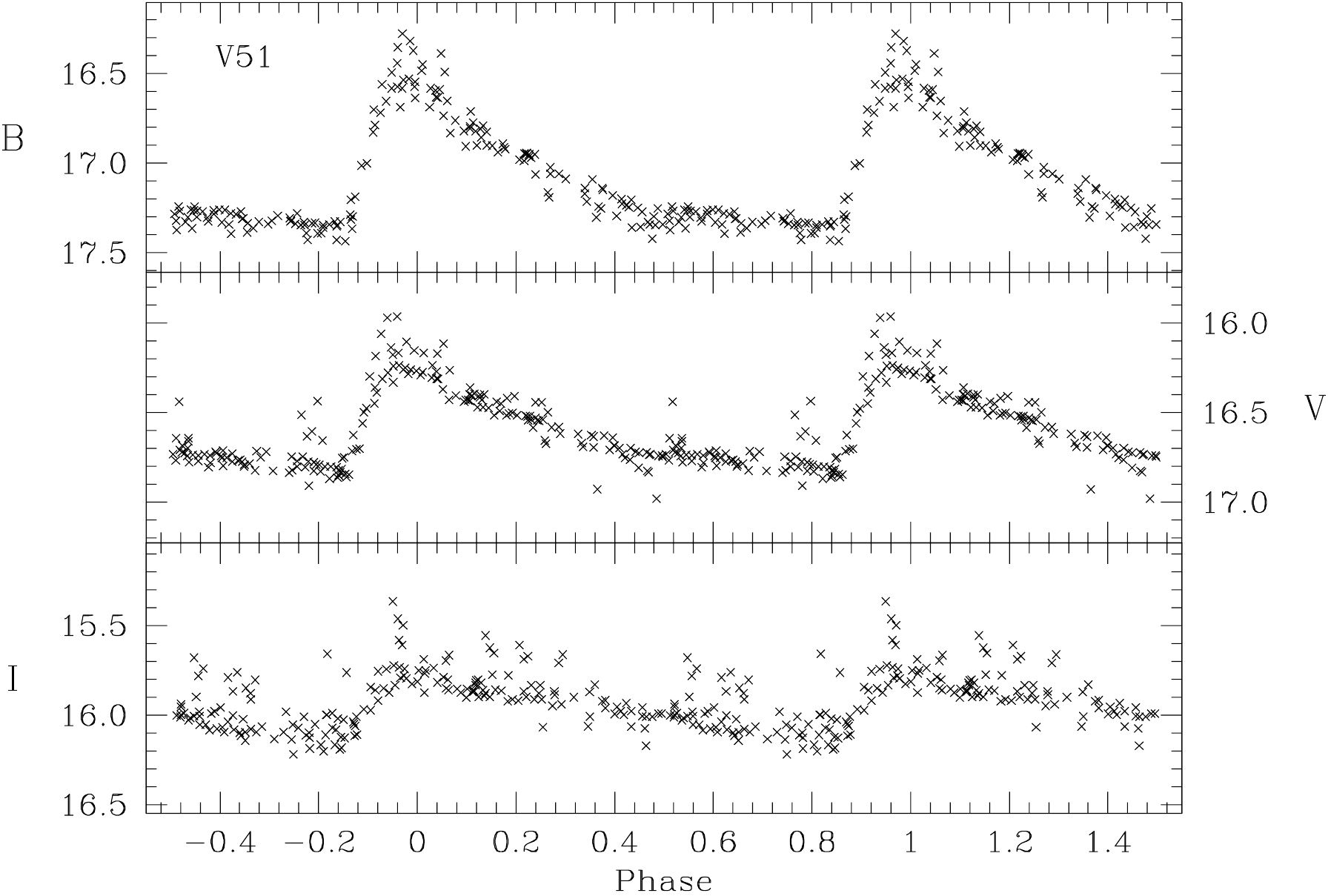}
    \includegraphics[width=.45\textwidth]{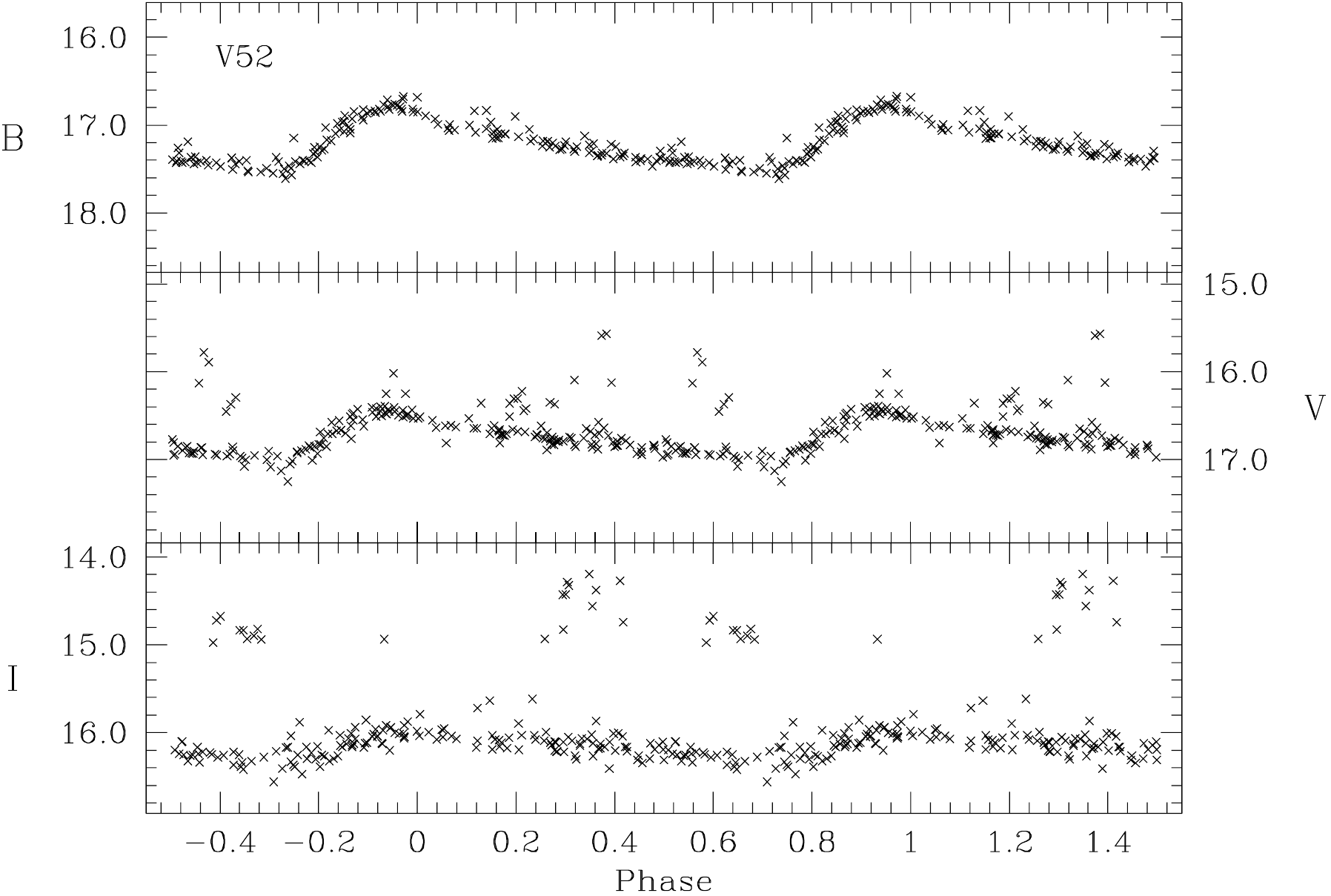}
    \includegraphics[width=.45\textwidth]{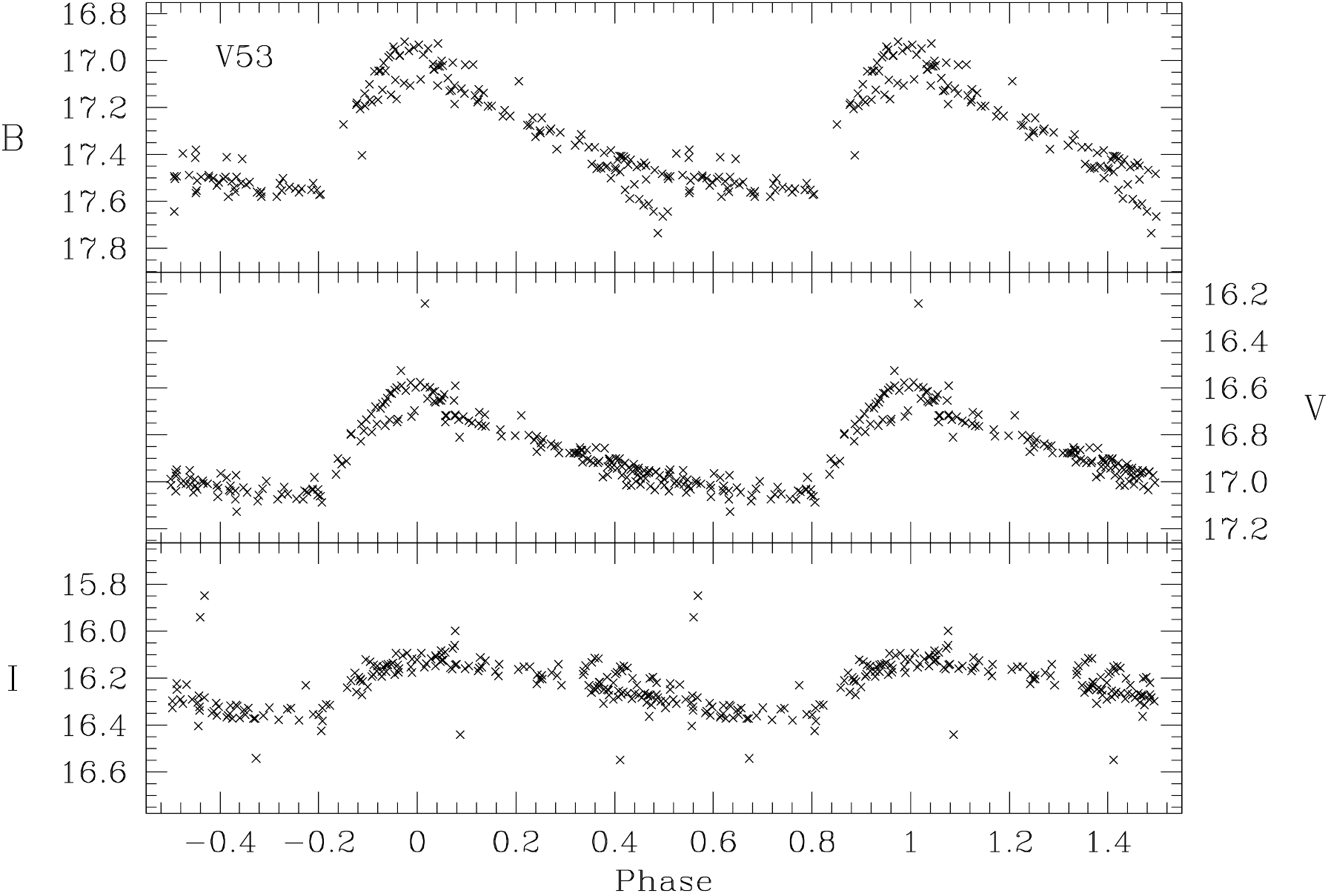}
     \caption{As in Figure~\ref{figvar1}, but for V45, V51, V52, and V53.}
    \label{figvar2}
    \end{figure}
   
  
   \begin{figure}
   \centering
    \includegraphics[width=.45\textwidth]{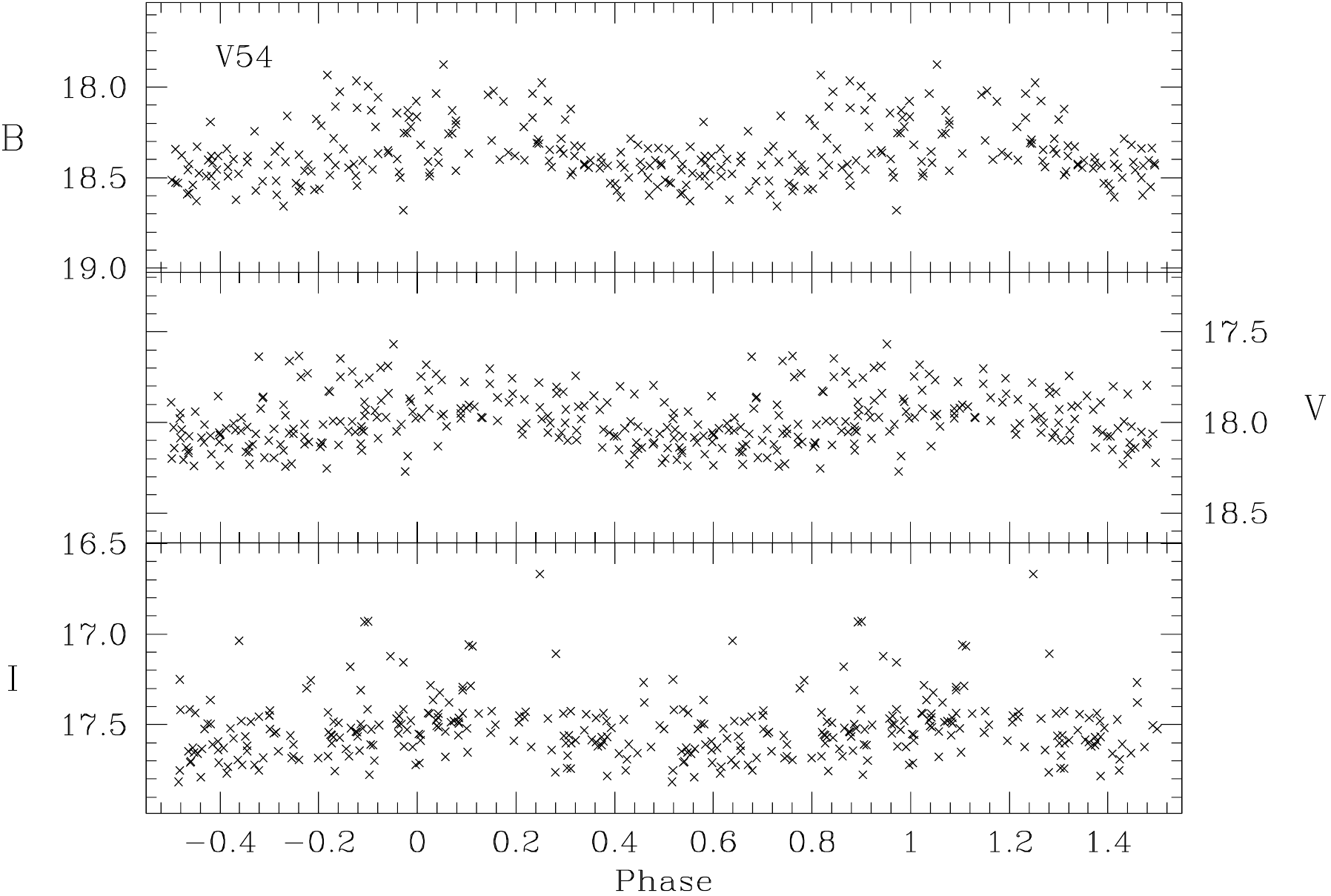}
    \includegraphics[width=.45\textwidth]{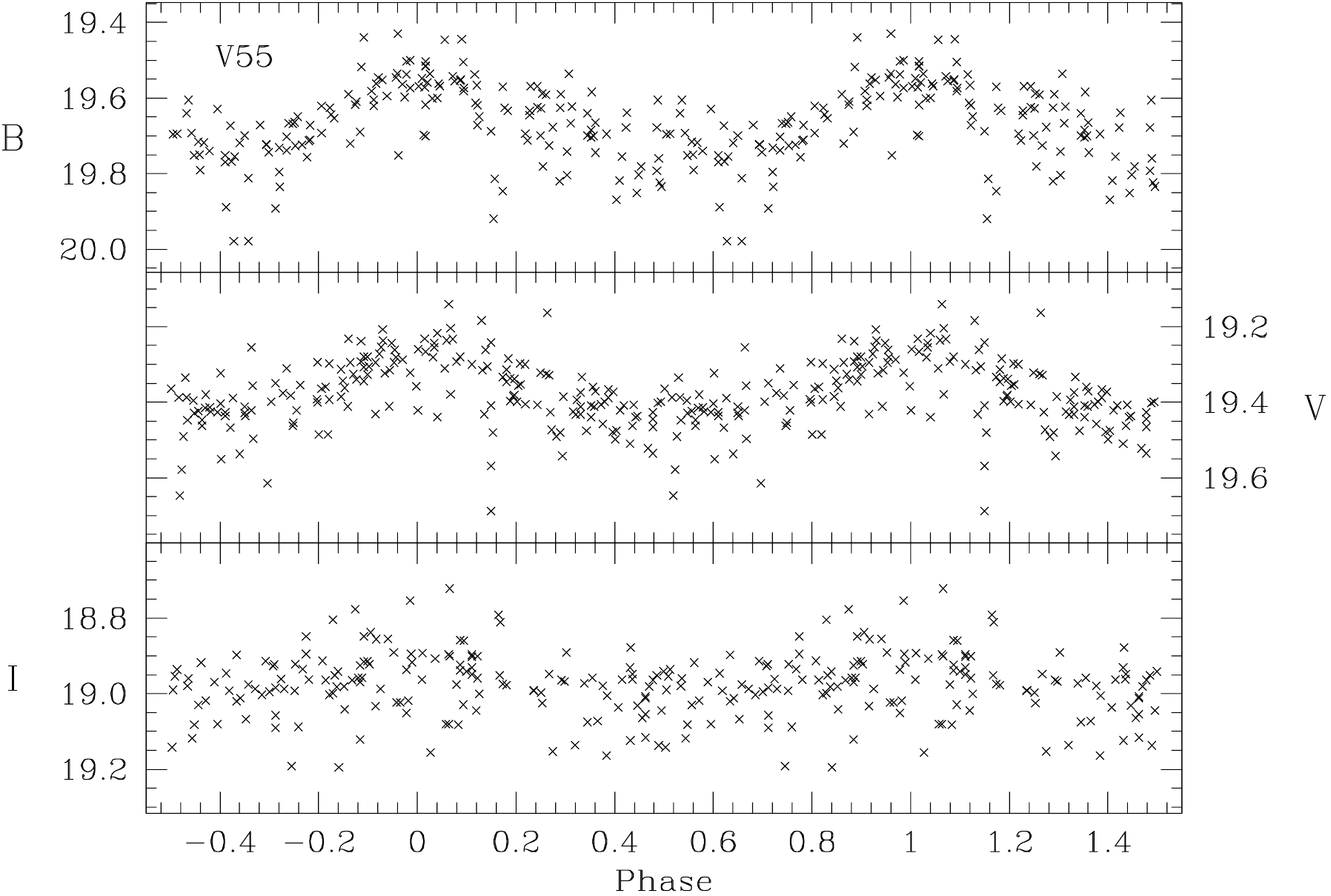}
    \caption{As in Figure~\ref{figvar1}, but for the SX Phe stars V54-V55.}
   \label{figSXPHE}
  \end{figure}
   
   
    \begin{figure}
   \centering
    \includegraphics[width=.45\textwidth]{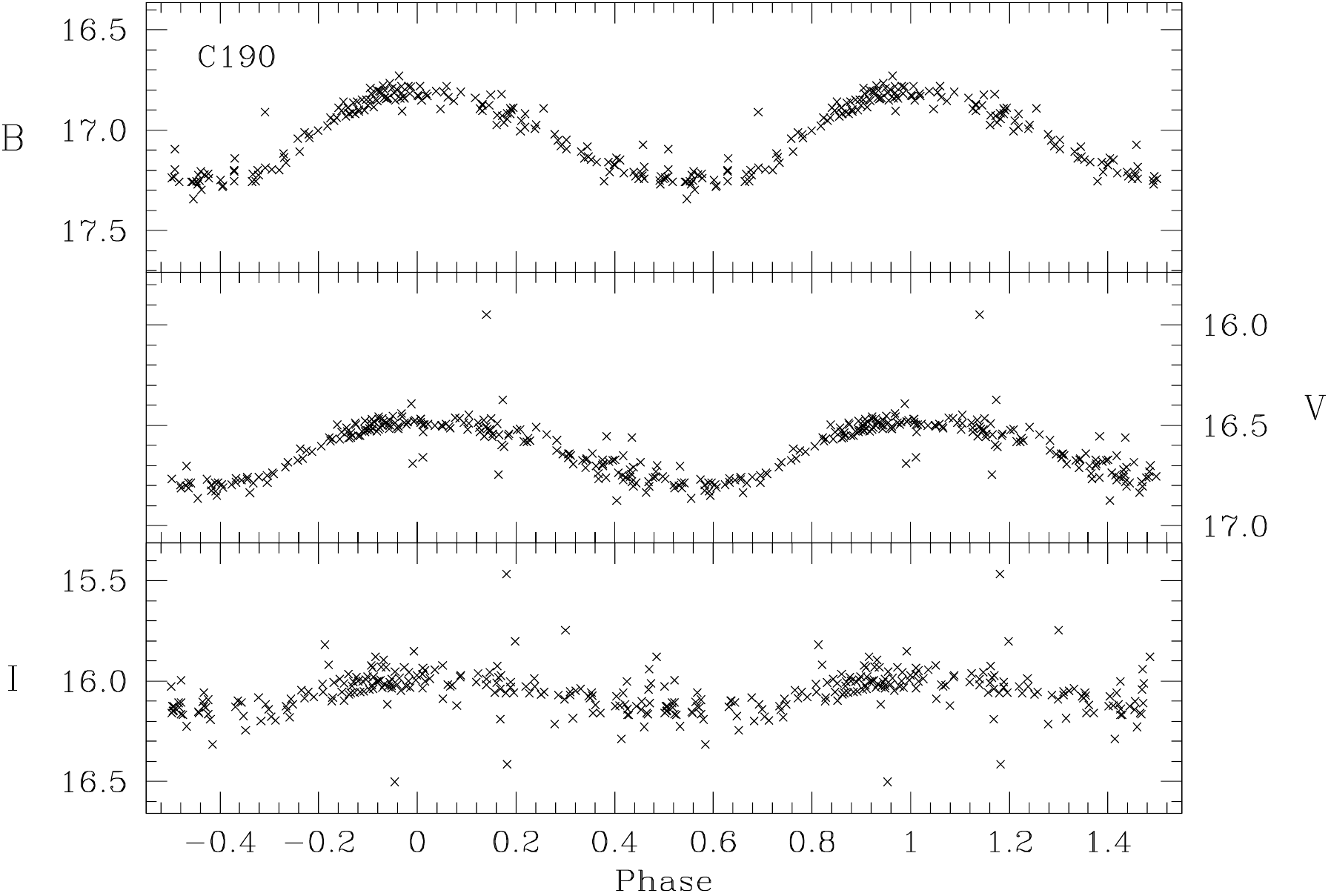}
    \includegraphics[width=.45\textwidth]{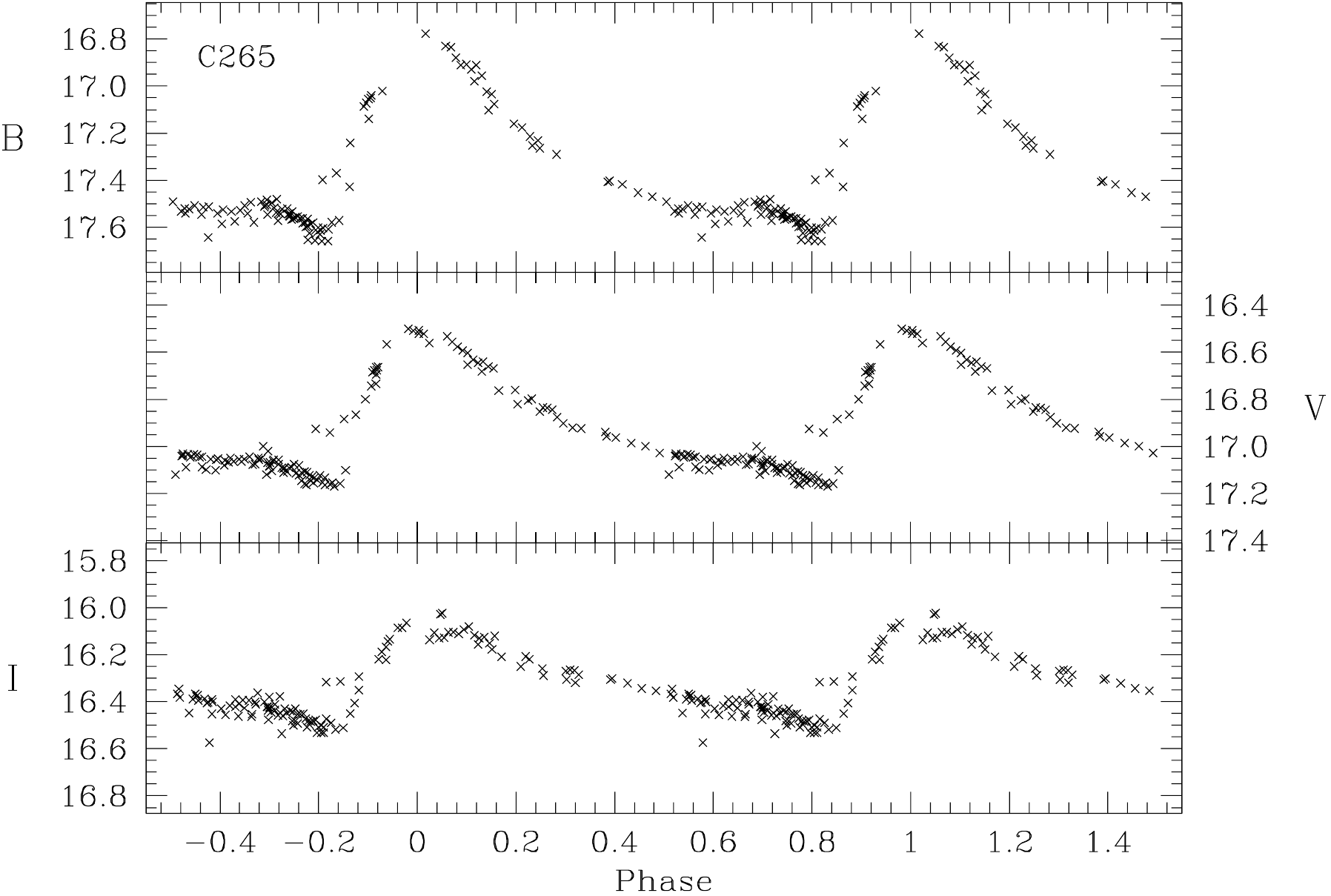}
   \includegraphics[width=.45\textwidth]{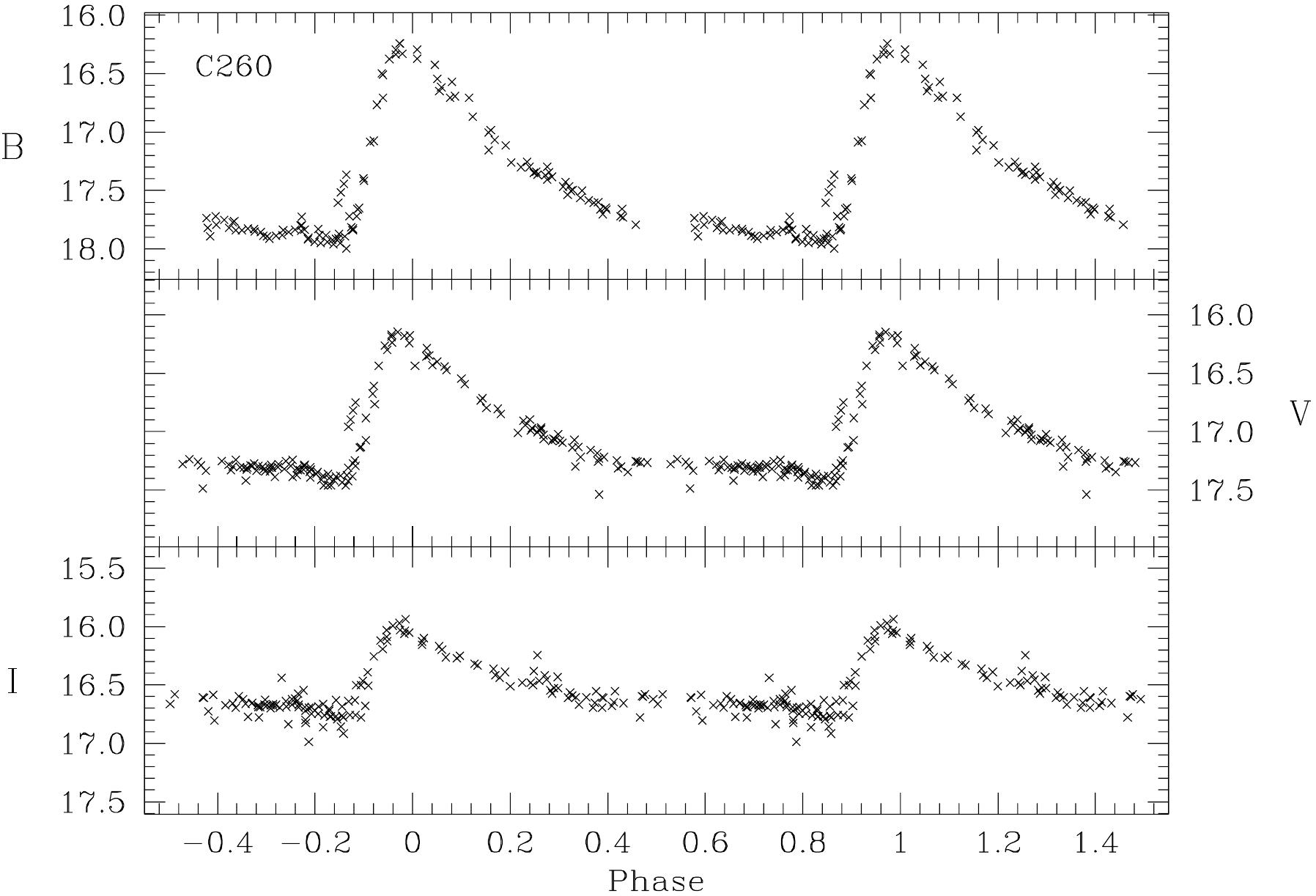}  
  \caption{As in Figure~\ref{figvar1}, but for the newly discovered RR Lyrae stars (from top) V57, V59 and V60}.
   \label{fignewvar}
   \end{figure}

The difference in magnitude between the HB and TO level are related with the age according to the following relation:

\begin{equation}
 \Delta {\log} \, t_9 = (0.44 + 0.04 \rm [Fe/H])\Delta 
\end{equation}

\noindent \citep{buoea93}, where 
$\Delta = \left(\Delta V_{\rm TO}^{\rm HB}\right)_{\rm GC1} - \left(\Delta V_{\rm TO}^{\rm HB}\right)_{\rm GC2}$, 
and $t_9$ is the age in Gyr.  For M3 we obtained
$\Delta V_{\rm TO}^{\rm HB} = 3.44$, whereas for \objectname{NGC~6981} we found $\Delta V_{\rm TO}^{\rm HB} = 3.42$. This leads to an age 
difference of $\Delta {\log} \, t_9 \sim 0.008$, which translates to about 0.2~Gyr, around 12~Gyr 
(using $\Delta t_9 = t_9 \, \ln (10) \, \Delta {\log} \, t_9$), 
in the sense that M3 is formally older~-- but essentially implying that M3 and \objectname{NGC~6981} 
are coeval, to within the errors. 

Another way to approach this issue is to compare directly the M3 and M72 CMDs. This is done in Figure~\ref{compM3}, where the ridgeline obtained for 
\objectname{NGC~6981} is overplotted on the M3 photometry. 
Both the data and the ridgeline have been dereddened, using the $E(\bv)$ values from the Dec. 2010 version of the Harris (1996)
catalog. The ridgeline of \objectname{NGC~6981} has been shifted 1.3~mag brighter, to account for the difference in distance 
modulus with respect to M3. The RGBs of the two clusters match reasonably well, confirming that they have similar 
metallicities Also, the match in luminosity of their subgiant branches confirms that the ages of the two clusters 
are basically indistinguishable.


 \section{Variable Stars Revisited}\label{sec:variables}

Bramich et al. (2011, hereafter B11) published a revised catalog of variable stars in \objectname{NGC~6981}, based on $V$-, $R$-,
and $I$-band photometry. This was the first variability study for this cluster in $\sim 40$ years, using CCD imaging and differential
image analysis techniques . It allowed the discovery of 14 new variable stars, leading to a total census of 43 variable stars, 
including 40 RR Lyrae and 3 SX Phoenicis (see their Table~2).

With the advantage of having a longer time baseline, we have revisited the variability in \objectname{NGC~6981}. We discovered 3 new RR~Lyrae
stars, including two fundamental-mode (RRab) and 1 first-overtone (RRc) pulsators. Following the B11 notation, we catalog these stars as 
V57, V59 and V60 (V58 is discuss in the text below). The remainder of the variables found in B11 were recovered (including the SX Phe stars),
and in some cases, we have improved the phased light curve and the period determination. Concerning the Blazhko (1907) effect, 
which is characterized by a long-term modulation of the light curve amplitude and shape, 
we confirmed the presence of the effect, claimed by B11, in V11, V14, V15, V23, V28, V31, V32, V36, V49; the effect was also present
in V35, V39, V51 and V53. Although V10 was also claimed to present Blazhko effect, we found no evidence in the phased light curve.
The coordinates, periods and variability types are provided in Table~\ref{tab:variables}, and Figure~\ref{CMDvar} displays the positions
of the variables in the CMD. In some cases, B11 could not study stars with previous claims of variability, due to technical limitations.
We provide light curves for these stars in Figures~\ref{figvar1} through \ref{fignewvar}, with comments on a few individual stars given
in the next section.

After this paper had been submitted, Skottfelt et al. (2013), announced the discovery of a new variable star in M72, namely V57. 
The classification and period of the variable matches our findings, but there is a discrepancy in the star's coordinates (they provide
$\rm \alpha=20^h53^m27\fs12$, $\rm \delta=-12\degr32\arcmin13.9\arcsec$). Moreover, they report another (unclassified) variable, V58, with a 
period of 0.285~d, in a position near V57 in our catalog ($\rm \alpha=20^h53^m27\fs38$, $\rm \delta=-12\degr32\arcmin13.3\arcsec$). The reason
for this discrepancy is unknown; however, we note that the coordinates provided in our work match reasonably well those provided by B11. 
To avoid confusion regarding the variable census, we kept their catalog identification for this variable.

\subsection{Comments on Individual Stars}\label{sec:commentvar}
\textbf{V27 and V35:}  These variables were detected by Dickens \& Flinn (1972), being originally classified as ab-type RR Lyrae; however, 
in the study of B11 they appeared out of their FOV, so even though they include them in their final catalog, no new periods could be
 determined. We confirmed the RRab type for V27 and V35, with a period of 0.673871~d and 0.543749~d, respectively (Fig.~\ref{figvar1}),
 consistent with the periods listed by Dickens \& Flinn.  We also note that V35 displays the Blazhko (1907) effect.

\textbf{V38 and V42:}  According to B11, Sawyer (1953) and Dickens \& Flinn
(1972) discovered that these stars were variable, but they were
unable to provide a light curve because V38 is in the proximity of a saturated star, and V42 was itself
saturated. In our analysis, V38 shows no sign of variability. In the case of V42, the light curve was phased with a period of 1.01109~d, as  
shown in Figure~\ref{V42}. In the CMD, the star is located at the red extension of the bright RGB, with a position consistent with that 
expected for the RGB tip, if the star is a bona-fide cluster member. 
Such short periods for bright RGB stars are quite unusual. The proximity of the period to 1~d renders the derived 
period suspicious. On the other hand, barring saturation effects, the variability amplitude, of order 0.5~mag in 
$B$, 0.3~mag in $V$, and 0.2~mag in $I$, seems quite significant. 

\textbf{V39:} This star shows no variability in B11, mostly due to outlier photometric measurements. In this study,
we found evidence of variability, and its light curve was phased with a period of 0.426785~d. The morphology of the light curve
suggests a RRab classification. The Blazhko (1907) effect might also be present. However we note that the star is fainter than the bulk of 
the cluster's RR Lyrae population, with $V = 17.95$. (see Fig.~13). Thus, V39 may be a field variable, and we have accordingly not included 
this star in our calculations of reddening and HB morphology parameters. 

\textbf{V44, V45, V52 and V53:}  B11 do not provide any period for these stars, since each of them suffered from poor phase coverage, 
due to the closeness of a saturated star affecting the light curve. We present here the improved light curves in Figures~\ref{figvar1} and
\ref{figvar2}, and their derived periods are given in Table~\ref{tab:variables}.

\textbf{V51:} B11 listed a period of 0.357335~d, although they noted that the period is not reliable due to poor phase coverage
and blending. In fact, the light curve they provided does not show the maximum. Here we provide a more reliable period
of 0.548599~d, and our well-sampled light curve is shown in Figure~\ref{figvar2}.

\subsection{SX Phoenicis Stars}\label{sec:SXP}
The number of SX Phe stars in GCs has increased markedly in recent years. These variables are of particular interest because they are 
located in the BSS zone in the CMD. Recently, Cohen \& Sarajedini (2012) compiled an updated SX Phe stars catalog in GCs,
establishing a period-luminosity relation for the sample. In the future, it is hoped that the pulsation properties of SX Phe
stars will help understand BSS formation and evolution.

In our study, we have recovered all 3 SX Phe stars found in B11. Using Period04 (Lenz \& Breger 2005) to analyze the Fourier
diagrams, the periods found correspond to the frequency of the largest amplitude oscillation, and the pulsation mode could not be identified
in two cases, V55 and V56, with $V$ amplitudes of 0.22 and 0.08~mag, respectively. Since there are no non-radial pulsators with $A_V > 0.15$
(Cohen \& Sarajedini 2012), V55 is likely a radial pulsator. However, the argument does not go in the other direction: double radial-mode
pulsators may have $A_V < 0.15$, and even some fundamental-mode SX Phe may have small amplitudes. 
In the case of V54, we detected two significant frequencies, with a ratio of  $f_1/f_2 \approx 0.78$, which are likely to correspond to the
fundamental mode ($f_2$) and the first radial overtone ($f_1$; Olech et al. 2005). This result is consistent with what B11 found for these
stars. 
The periods of V54 and V56 are slightly different from B11, while for V55 we obtained a good agreement (see Figures~\ref{figSXPHE} for the
phased light curves). It should be noted that in the case of V56 we could not phase the light curve convincingly enough,
and the period may not be reliable. The amplitude found by Period04, $V \approx$~0.08, is comparable with the error in the photometry. 
To close, we note that V54 appears to be slightly brighter than the limits of the BSS region, as defined by B11 (based on Harris 1993).

\section{Summary}\label{sec:conclu}
We present the most comprehensive study of the CMD and variable star content of the GGC \objectname{M72} to date, based on new and
archival $BVI$ CCD images. Our CMD reaches almost four magnitudes below the turn-off level, which allowed us to conclude that M72 
has the same age as M3 (which has a similar metallicity and HB morphology), 
to within the errors. Based on the measured RGB color and slope, we infer that the cluster has a metallicity 
${\rm [Fe/H]} \simeq -1.50$ in the new UVES scale. We firmly detect the cluster's blue straggler population, which is found to be 
more centrally concentrated than the RGB component. We also find evidence of extratidal cluster stars being present out to 
$r \approx 14.1\arcmin$, or about twice M72's tidal radius, and speculate that tidal tails associated with the 
cluster may exist. Finally, we revisit the variable star content of the cluster, recovering all previous known variables, including 
three SX Phe stars, and discovering three previously unknown RR Lyrae (1 c-type and 2 ab-type).

\acknowledgments 
We thank Alistair Walker for his help with the database
and the anonymous referee for some comments that helped improve this manuscript. 
This project is supported by the Chilean Ministry for the
Economy, Development, and Tourism's Programa Iniciativa Cient\'{i}fica 
Milenio through grant P07-021-F, awarded to The Milky Way Millennium 
Nucleus; by the BASAL Center for Astrophysics and Associated Technologies 
(PFB-06); by Proyecto Fondecyt Regular \#1110326; and by Proyecto Anillo 
ACT-86. PA acknowledges the support by ALMA-CONICYT project \#31110002. 
MZ acknowledges support by Fondecyt Regular 1110393 and by the John
Simon Guggenheim Memorial Foundation Fellowship. HAS thanks the U.S. National
Science Foundation for support under grants AST 0607249 and 0707756.

\end{document}